\documentclass{emulateapj}

\shorttitle{Re-appearance of McNeil's nebula (V1647 Orionis)}
\shortauthors{Ninan et al.}

\begin{document}
\title{Re-appearance of McNeil's nebula (V1647 Orionis) and its outburst environment}

\author{J. P. Ninan, D. K. Ojha}
\affil{Department of Astronomy and Astrophysics, Tata Institute of Fundamental Research, 
Homi Bhabha Road, Colaba, Mumbai 400 005, India}
\email{ninan@tifr.res.in}
\author{B. C. Bhatt}
\affil{Indian Institute of Astrophysics, Koramangala, Bangalore 560 034, India}
\author{S. K. Ghosh}
\affil{National Centre for Radio Astrophysics, Tata Institute of Fundamental Research, 
Pune 411 007, India}
\author{V. Mohan}
\affil{Inter-University Centre for Astronomy and Astrophysics, Pune 411 007, India}
\author{K. K. Mallick}
\affil{Department of Astronomy and Astrophysics, Tata Institute of Fundamental Research, 
Homi Bhabha Road, Colaba, Mumbai 400 005, India}
\author{M. Tamura}
\affil{National Astronomical Observatory of Japan, Mitaka, Tokyo 181-8588, Japan}
\and
\author{Th. Henning}
\affil{Max-Planck-Institute for Astronomy, K\"onigstuhl 17, 69117 Heidelberg, Germany}

\begin{abstract}
We present a detailed study of McNeil's nebula (V1647 Ori) in its ongoing outburst phase starting from September 2008 to March 2013. Our 124 nights of photometric observations were carried out in optical \textit{V}, \textit{R}, \textit{I} and near-infrared (NIR) \textit{J}, \textit{H}, \textit{K} bands, and 59 nights of medium resolution spectroscopic observations were done in 5200 - 9000 $\mathring{A}$ wavelength range. All observations were carried out with 2-m Himalayan \textit{Chandra} Telescope (HCT) and 2-m IUCAA Girawali Telescope. Our observations show that over last four and a half years, V1647 Ori and the region C near Herbig-Haro object, HH 22A, have been undergoing a slow dimming at a rate of $\sim 0.04$ mag yr$^{-1}$ and $\sim 0.05$ mag yr$^{-1}$ respectively in \textit{R}-band, which is 6 times slower than the rate during similar stage of V1647 Ori in 2003 outburst. We detected change in flux distribution over the reflection nebula implying changes in circumstellar matter distribution between 2003 and 2008 outbursts. Apart from steady wind of velocity $\sim 350$ km s$^{-1}$ we detected two episodic magnetic reconnection driven winds. Forbidden [O I] 6300$\mathring{A}$ and [Fe II] 7155$\mathring{A}$ lines were also detected implying shock regions probably from jets. We tried to explain the outburst timescales of V1647 Ori using the standard models of FUors kind of outburst and found that pure thermal instability models like \citet{bell94} cannot explain the variations in timescales. In the framework of various instability models we conclude that one possible reason for sudden ending of 2003 outburst in 2005 November was due to a low density region or gap in the inner region ($\sim$ 1 AU) of the disc.
\end{abstract}


\section{Introduction}
When low-mass stars like our Sun are born, they slowly accrete gas from the collapsing cloud through an accretion disc. Collimated outflows are also typically seen in most of these objects. 
The discontinuous knots seen in outflows from these objects, and the mismatch of accretion rate between envelope to disc and disc to star (known as \textquotedblleft Luminosity Problem\textquotedblright) in young stellar objects (YSOs), all hint to an episodic nature of  accretion instead of an ideal steady inflow \citep{kenyon90,evans09,ioannidis12}.
The other important feature seen in these objects is the outburst.
Rare outbursts that we see among these YSOs 
are found to be correlated with an order of magnitude increase in mass infall rate and they could be the episodic accretion events required to explain the outflow discontinuities and \textquotedblleft Luminosity Problem\textquotedblright. These outbursts are empirically classified as FUors (decades long outbursts with 4-5 magnitude change in optical) and EXors (few months-years long outbursts with 2-3 magnitude change in optical) \citep{herbig77,hartmann98,hartmann96}. Due to the short timescales of outbursts in comparison to millions of years timescale of star formation, these events are extremely rare and only less than a dozen confirmed FUor outbursts have been detected so far. From their frequency of outburst it is estimated that every low-mass YSO should go through these outbursts at least 50 times in its protostellar phase \citep{scholz13}. Regarding the discontinuous knots seen in outflows, it should be noted that the timescales between discontinuities were estimated to be of the order of $10^{3}$ years by \citet{ioannidis12} and the timescales between FUor outbursts in a single star were estimated to be of the order of $10^{4}$ years by \citet{scholz13}. So the discontinuities could be due to some other shorter timescale variations in accretion rate rather than classical FUors. But as \citet{scholz13} pointed out, we do not have a good estimate of FUors' timescales in the early Class I stage. Hence we cannot rule out the possibility that discontinuities in outflows are created during FUors events.

One such object which was extensively studied recently in literature is \object{V1647 Ori} (V1647 Orionis), 400 pc away in the L1630 dark cloud of Orion. It underwent a sudden outburst of $\sim 5$ mag in optical in 2003 \citep{mcneil04,briceno04} and illuminated a reflection nebula, now named as McNeil's nebula after its discoverer Jay McNeil. 
\citet{reipurth04} reported 3 mag increase in near-infrared (NIR) and \citet{andrews04} reported 25 times increase in 12 $\mu m$ flux and no flux change in submillimeter of V1647 Ori during the outburst. \citet{kastner04} reported a 50 times increase in X-ray flux of V1647 Ori during the outburst, and noted that the derived temperature of the plasma is too high for accretion alone to explain and hinted that magnetic reconnection events might be heating up the plasma. Based on the upper limit on radio continuum emission from McNeil's nebula at 1272 MHz from Giant Metrewave Radio Telescope (GMRT), India, \citet{vig05} constrained the extent of HII region corresponding to a temperature $\gtrsim$ 2500 K to be $\lesssim$ 26 AU. Spectroscopic studies showed strong H$\alpha$ and Ca II IR triplet lines in emission \citep{briceno04, ojha06}. In NIR region, strong CO bandheads (2.29 $\mu m$) and Br$\gamma$ line (2.16 $\mu m$; implying strong accretion) were detected in emission \citep{reipurth04,vacca04}. The strong P-Cygni profile in H$\alpha$ emission indicated wind-velocity ranging from 600 - 300 km s$^{-1}$ \citep{ojha06,vacca04}. \citet{abraham06} carried out AU scale observations using Very Large Telescope Interferometer/ Mid-Infrared interferometric Instrument (VLTI/MIDI). By fitting both spectral energy distribution (SED) and visibility values they deduced a moderately flaring disc with temperature profile $T \sim r^{-0.53}$ (T(1AU)=680K) and mass $\sim 0.05 M_{\odot}$, with inner radius of 7 $R_{\odot}$ (0.03 AU) and outer radius of 100 AU. This temperature profile is shallower than the $T \sim r^{-0.75}$ canonical model \citep{pringle81}. They also reported that the mid-infrared emitting region at 10 $\mu m$ has a size of $\sim$ 7 AU. \citet{rettig05} used the CO lines in infrared (IR) to measure the temperature of the inner accretion disc region which was estimated to be T $\approx$ 2500 K. 

\citet{ojha06} and \citet{kospal05} reported a sudden dimming and termination of the 2003 outburst in November 2005. Thus, 2003 outburst lasted for a total of $\sim$ 2 years and V1647 Ori returned to its pre-outburst phase in early 2006.
\citet{acosta07} estimated the inclination angle of disc to be $61^\circ$ and also estimated the accretion rate to be $5 \times10^{-6} M_{\odot}$ yr$^{-1}$ during outburst and  $5\times 10^{-7} M_{\odot}$ yr$^{-1}$ in 2006 quiescent state.
\citet{aspin06} reported that $\sim$ 37 years prior to 2003 outburst, i.e. in 1966, V1647 Ori had undergone a similar magnitude of outburst, lasting for a duration somewhere between 5 to 20 months.

\begin{figure}
\begin{center}
\includegraphics[width=1\textwidth]{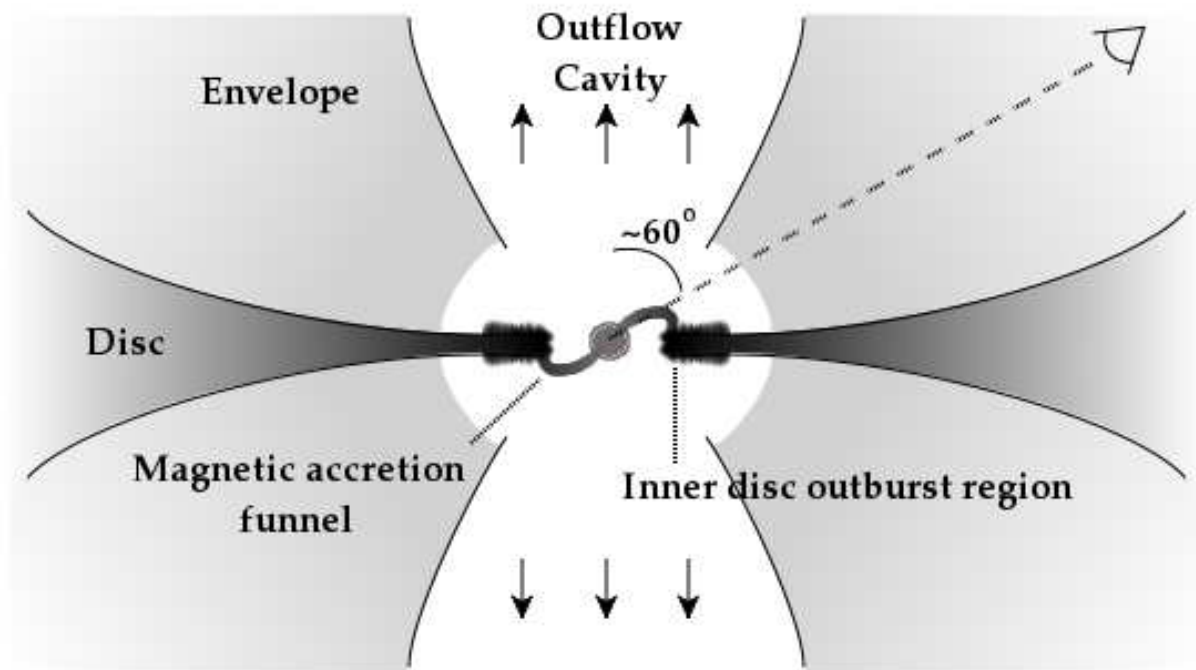}%{CrossSection_of_V1647Disc.eps}
\caption{Cross-section of V1647 Ori surroundings we know so far and the line of sight angle to the disc. Image is not drawn to scale.}
\label{img:crossSec-Diag}
\end{center}
\end{figure}

Contrary to expected decades long quiescense, \object{V1647 Ori} underwent a second outburst in 2008 just after spending two years in quiescent state \citep{aspin09}. It brightened up to the same magnitude and had almost identical spectral features in optical and NIR as the first outburst. One striking difference was that the strong CO bandhead emission at 2.29 $\mu m$ was absent in second outburst \citep{aspin11}. The X-ray flux with plasma temperature of 2-6 keV during both outbursts was postulated to be due to magnetic reconnection events in the disc-star magnetic field interaction \citep[and references therein]{teets11}. \citet{hamaguchi12} normalised and combined both outbursts' data in X-ray and detected one day periodicity in light curve, which they modeled with two accretion hot spots on the top and bottom hemispheres of the star rotating with one day period and inclination of $68^\circ$. Figure \ref{img:crossSec-Diag} shows the overall cross-section picture of the surroundings of V1647 Ori we know so far.

V1647 Ori provides a unique opportunity to understand the physical processes undergone in FUors or EXors kind of outbursts. The short time scale behaviors of this object make it possible for us to make a detailed study of the object. In literature there exists mainly three kinds of model for explaining the outburst phenomena (see Section \ref{modelsNDiscuss}). 
The differences between these models are all in the inner region of disc ($<$ 1 AU), and optical and NIR are the right wavelength regime to probe this region of the disc.
Detailed understanding of V1647 Ori will thus provide us a laboratory to check our understanding of various instabilities like thermal, gravitational and magnetorotational in proto-planetary disc around young low-mass stars. % like our Sun.

We have carried out continuous observations for more than four and a half years (2008 - 2013) of V1647 Ori in optical and NIR wavelengths for detailed study of its dynamics during outburst and post-outburst stages of the second outburst. This data combined with previous outbursts' provide us more insight on the nature of outburst and also constrain the existing physical models.
In this paper, we present the results of our long-term optical and NIR photometric and spectroscopic observations of the outburst source and associated McNeil's nebula. In Section \ref{obsNreduction} we describe the observational details and the data reduction procedures. In Section \ref{resultsNDiscuss} we present our new findings and results from observations.
In Section \ref{modelsNDiscuss} we analyse the ability of each existing physical models to explain V1647 Ori's outburst history.
Finally, in Section \ref{conclusions} we summarise our main results.

%-------------------------------------------------------------
\section{Observations and Data Reduction}
\label{obsNreduction}
\subsection{Optical Photometry}
%---------
Our long-term optical observations span from 2008 September 14 to 2013 March 11 and were carried out with 2-m Himalayan \textit{Chandra} Telescope (HCT) at Indian Astronomical Observatory, Hanle (Ladakh), India and with 2-m Inter-University Centre for Astronomy and Astrophysics (IUCAA) Girawali Telescope at IUCAA Girawali Observatory (IGO), Girawali (Pune), India. At HCT, for photometry central 2K $\times$ 2K section of Himalaya Faint Object Spectrograph \& Camera (HFOSC) CCD, which has a pixel scale of 0.296 arcsec was used, giving us a field of view (FoV) of $\sim 10 \times 10$ arcmin$^{2}$. At IGO, 2K $\times$ 2K IUCAA Faint Object Spectrograph \& Camera (IFOSC) CCD was used which also has a similar pixel scale of 0.3 arcsec, giving us a FoV of $\sim 10 \times 10$ arcmin$^{2}$. 
Further details of the instruments and telescopes are available at \url{http://www.iiap.res.in/iao/hfosc.html} and \url{http://www.iucaa.ernet.in/~itp/igoweb/igo\_tele\_and\_inst.htm}. Out of our total observation of 110 nights, 84 nights were observed from HCT and 26 nights from IGO.
 
\begin{figure}
\begin{center}
\includegraphics[width=1\textwidth]{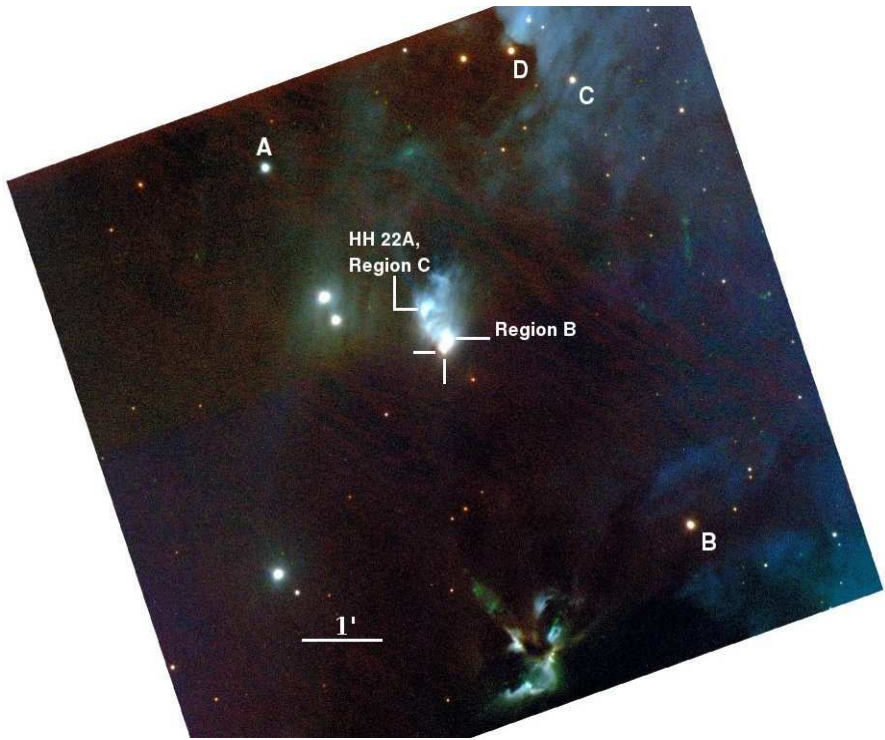}%{McNeil_RGB4paper.eps}
\caption{Color-composite image of McNeil's nebula (V1647 Ori) region taken from IGO (\textit{V}: blue, \textit{R}: green, \textit{I}: red) on 2010 February 13. FoV is $\sim 10 \times 10$ arcmin$^2$. North is up and east is to the left-hand side. Stars marked as A, B, C and D are the secondary standard stars used for magnitude calibration. The location of V1647 Ori is marked at the center by two perpendicular lines. Region C, overlapping Herbig-Haro object HH 22A, is marked together. A knot in south-western section (region B) of nebula is also marked.  At 400 pc distance, the scale 1\arcmin\,  corresponds to 24000 AU.}
\label{img:field}
\end{center}
\end{figure}

The V1647 Ori's field $(\alpha,\delta)_{2000} = (05^h46^m13^s.135, -00^{\circ}06^{'}04^{''}.82)$  was observed in standard $VRI$ Bessel filters. Nearby Landolt's standard star fields \citep{landolt92} were also observed for magnitude calibration and for solving color transformation equation coefficients of each night. For nights which do not have standard star observations, we identified six stars in the object's frame whose magnitudes remain constant throughout. Four of them were used as secondary standards (see Figure \ref{img:field}) and other two were used to check consistency and error. Apart from object frames, bias and sky flats were also taken in each night for the basic data reduction. For fringe removal in IGO \textit{I}-band images, blank sky frames were also taken. 
The log of photometric observations is given in Table \ref{table:Obs_Log}. Only a portion of the table is provided here. The complete table is available in electronic form as part of the online material.

Blank sky images in \textit{I}-band were used to create fringe templates by MKFRINGECOR task in IRAF\footnote{IRAF is distributed by the National Optical Astronomy Observatories, which are operated by the Association of Universities for Research in Astronomy, Inc., under contract with the National Science Foundation.}, which were later used to subtract the fringes that appeared in \textit{I}-band images taken from IGO.  Data reduction was done with the semi-automatic pipeline written in PyRAF\footnote{PyRAF and PyFITS are products of the Space Telescope Science Institute, which is operated by AURA for NASA} and IRAF CL scripts.
Standard photometric data reduction steps like bias-subtraction and median flat-fielding were done for all the nights. 
Point-spread function (PSF) photometry (using PSF \& ALLSTAR tasks in DAOPHOT package of IRAF) on V1647 Ori was not able to fully remove the nebular contamination. We found a strong correlation between fluctuation in magnitude of V1647 Ori and fluctuation in atmospheric seeing condition. 
This is because the contamination of flux from nebula into V1647 Ori's aperture
was a function of atmospheric seeing. So we generated a set of images by convolving each frame with 2-D Gaussian kernel of different standard deviation (using IMFILTER.GAUSS task in IRAF) for simulating different atmospheric seeing conditions. We then recalculated the magnitudes by DAOPHOT, PSF and ALLSTAR algorithms of IRAF for various atmospheric seeing conditions. The differential magnitudes obtained from each frame's set was interpolated to obtain magnitude at an atmospheric seeing of 1.18 arcsec, which was taken to be the seeing to be interpolated to, for all nights and it was chosen to minimise interpolation error. This method reduced our error bars in magnitude by a factor of 2. Apart from the Gaussian convolution step, the PSF photometry steps were all same as \citet{ojha06}.

Magnitudes of the whole nebula and other objects in the nebula like region C (near HH 22A) and region B defined by \citet{briceno04} in their figure 2 (also see Figure \ref{img:field}) were measured by simple aperture photometry with an aperture radius of 80 arcsec for nebula and 12 arcsec for the regions C and B. For obtaining the flux, 
the aperture of the objects like regions C and B were centered at the objects itself.

\subsection{Optical Spectroscopy}
%----------------------
Our long-term spectroscopic observations also span the same duration as that of photometric observations (2008 September to 2013 March) using both 2-m HCT and 2-m IGO. The full 2K $\times$ 4K section of HFOSC CCD spectrograph  was used in HCT observations and 2K $\times$ 2K IFOSC CCD spectrograph was used for IGO observations. Spectroscopic observations were carried out on 35 nights from HCT and 24 nights from IGO, thus totalling to 59 nights of V1647 Ori's spectroscopic observations.
The log of spectroscopic observations is listed in Table \ref{table:Obs_Log}. Only a portion of the table is provided here. The complete table is available in electronic form as part of the online material. 
In order to detect the prominent $H\alpha$ $\lambda6563$ $\mathring{A}$ and Ca II IR triplet lines ($\lambda8498$, $\lambda8542$, $\lambda8662$ $\mathring{A}$)%$\lambda8542$ $\mathring{A}$ 
, we observed in the effective wavelength range of $5200 - 9000$  $\mathring{A}$ using grism 8 (center wavelength 7200 $\mathring{A}$ ) and grism 7 (center wavelength 5300 $\mathring{A}$).
The spectral resolution obtained for grism 8 and 7 with 150 micron slit at IGO and 167 micron slit at HCT was $\sim 7$ $\mathring{A}$. Nebulosity contamination in spectrum of V1647 Ori was minimised by keeping the slit in east-west orientation.
Standard IRAF tasks like APALL and APSUM were used for spectral reduction. Wavelength calibration was carried out using the FeNe, FeAr and HeCu lamps. For final measurement of equivalent width the extracted 1-D spectra were normalised with respect to continuum. 
For spectroscopic data reduction of HCT and IGO data, semi-automated pipeline written in PyRAF was used.

\subsection{Near-Infrared Photometry}
Along with optical monitoring we also carried out photometric monitoring in \textit{JHK} bands using the HCT NIR camera (NIRCAM) and TIFR  NIR Imaging Camera-II (TIRCAM2). NIRCAM has a $512 \times 512$ Mercury Cadmium Telluride (HgCdTe) array, with a pixel size of 18 $\mu m$, which gives a FoV of $3.6 \times 3.6$ arcmin$^{2}$ with HCT. Filters used for observation were J ($\lambda_{center}$= 1.28 $\mu m$, $\Delta\lambda$= 0.28 $\mu m$), H ($\lambda_{center}$= 1.66 $\mu m$, $\Delta\lambda$= 0.33 $\mu m$) and K ($\lambda_{center}$= 2.22 $\mu m$, $\Delta\lambda$= 0.38 $\mu m$). Further details of the instrument are available at \url{http://www.iiap.res.in/iao/nir.html}.
TIRCAM2 has a $512 \times 512$ Indium Antimonide (InSb) array with a pixel size of 27 $\mu m$. We observed McNeil's nebula during the engineering run of TIRCAM2 at  2-m IGO and 1.2-m Physical Research Laboratory (PRL) Mount Abu telescope. Filters used for observation were J ($\lambda_{center}$= 1.20 $\mu m$, $\Delta\lambda$= 0.36 $\mu m$), H ($\lambda_{center}$= 1.66 $\mu m$, $\Delta\lambda$= 0.30 $\mu m$) and K ($\lambda_{center}$= 2.19 $\mu m$, $\Delta\lambda$= 0.40 $\mu m$). Further details of the instrument are available in \citet{naik12}. We have a total of 14 nights of NIR photometric observations, with the first set of data  taken  during the quiescent phase in 2007, i.e. before the 2008 outburst.
Observations of V1647 Ori were carried out by taking several sets of exposures; each set contains exposure with the telescope pointing at five different dithered positions. %close to the target position. 
The master sky frame for sky-subtraction was generated by median combining all the dithered object frames.
Data reduction and final photometry were done using standard IRAF aperture photometric tasks. To be consistent with magnitude estimates by \citet{ojha06}, for flux calibration we used an aperture $\sim 7$ arcsec, and for background sky estimation we used 
an annulus with an inner radius of $\sim$ 50\arcsec\, and width $\sim$ 5\arcsec\,. 
For instrumental to apparent magnitude calibration, we observed standard stars around AS13, AS9 and HD225023 fields \citep{hunt98} on the same night with similar airmass as V1647 Ori observations. On 2011 December 6, standard stars were not observed, hence we used the magnitude measured on other nights of the nearby star (2MASS J05461162-0006279) for photometric calibration.

%-----------------------

\section{Results and Discussion}
\label{resultsNDiscuss}
%-----------------------
\subsection{Photometric Results}
\begin{figure}
  \includegraphics[width=1\textwidth]{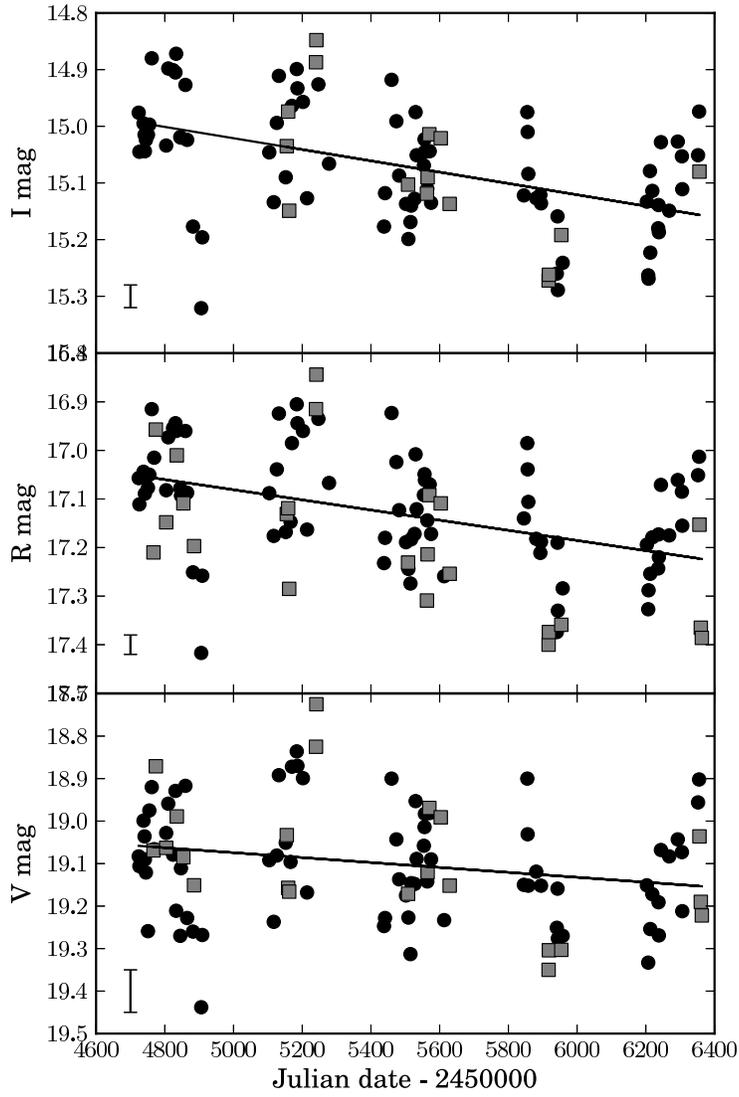}%{JD-v1647-mag_hct-igo.eps}
  \caption{Magnitude variation of V1647 Ori in the \textit{I, R} and \textit{V}-band from 
September 2008 to March 2013. Typical photometric error bar is given at the left bottom corner. The filled circles are from 
HCT and filled squares are from IGO measurements. The rate of dimming in \textit{I, R} and \textit{V}-bands are $0.036 \pm 0.007$ mag yr$^{-1}$, $0.038 \pm 0.007$ mag yr$^{-1}$ and $0.021 \pm 0.009$ mag yr$^{-1}$ respectively. }
\label{img:V1647Lightcurve}
\end{figure}

\begin{figure}
  \includegraphics[width=1\textwidth]{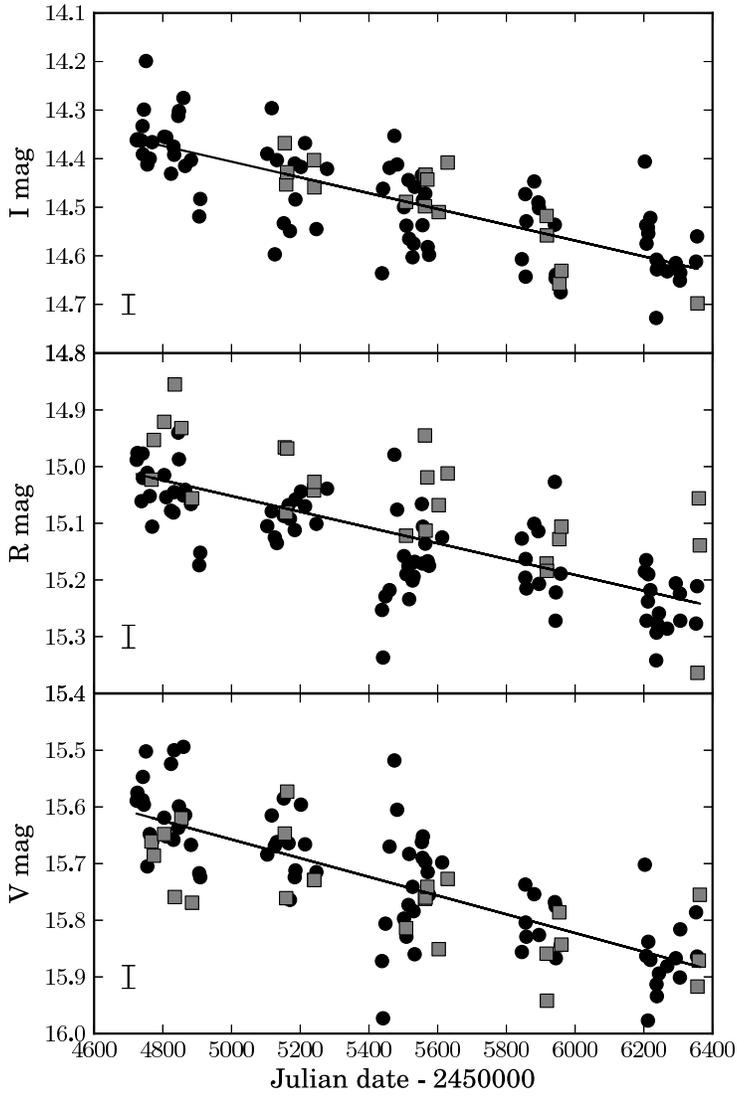}%{JD-HH22-mag_hct-igo.eps}
  \caption{Magnitude variation of region C, illuminated 
by V1647 Ori, in the \textit{I, R} and \textit{V}-band from September 2008 to March 2013. Typical photometric error bar is given at the left bottom corner. The filled circles are from 
HCT and filled squares are from IGO measurements. The rate of dimming in \textit{I, R} and \textit{V}-bands are $0.059 \pm 0.005$ mag yr$^{-1}$, $0.051 \pm 0.005$ mag yr$^{-1}$ and $0.060 \pm 0.005$ mag yr$^{-1}$ respectively.}
\label{img:HH22Lightcurve}
\end{figure}

%-------------
Figure \ref{img:field} shows the three-color composite image (\textit{V}: blue, \textit{R}: green, \textit{I}: red) of the McNeil's nebula field (FoV $\sim 10 \times 10$ arcmin$^{2}$) obtained from IGO on 2010 February 13. Secondary standard stars used for flux calibration are marked as A, B, C and D. The outburst source V1647 Ori, illuminating the nebula, is marked at the center. The region C, possibly unrelated to Herbig-Haro object, HH 22A, which is illuminated by V1647 Ori is also marked.
V1647 Ori had already reached its peak outburst phase before our first optical observation in September 2008. Its light curve steadily continued in peak outburst flux (\textquotedblleft high plateau\textquotedblright) phase even until our last observation taken in March 2013. 
However, our long-term continuous monitoring from 2008 September 14 to 2013 March 11 shows a slow but steady linear declining trend in the brightness of the source and nebula \citep{ninan12}. The linear slopes and the error in estimates of slopes were obtained by simple linear regression by ordinary least square fitting.
\textit{V}, \textit{R} and \textit{I} magnitudes of V1647 Ori 
and of region C, which is illuminated by the V1647 Ori from its face-on angle of the disc, are listed in Table \ref{table:mags}. Only a portion of the table is provided here. The complete table is available in electronic form as part of the online material. 
Light curves of V1647 Ori in \textit{I} and \textit{R}-bands clearly show a steady dimming (see Figure \ref{img:V1647Lightcurve}). During the last four and a half years of its second outburst, the brightness in \textit{I} and \textit{R} bands have decreased by $\sim 0.2$ mag. The rate of decline in magnitude of V1647 Ori is 0.036$\pm 0.007$ mag yr$^{-1}$ in \textit{I}-band and 0.038$\pm 0.007$ mag yr$^{-1}$ in \textit{R}-band. We do not see any statistically significant decline in \textit{V}-band magnitude of V1647 Ori. This could be due to higher fraction of contamination of nebula over V1647 Ori's aperture and slightly higher error in magnitudes due to faintness of source in \textit{V}-band. These flux changes are along our direct line of sight at an angle of $\sim 30^\circ$ to the plane of disc \citep{acosta07}. However, the flux measured along the cavity in perpendicular direction to the disc, which is reflected from region C, shows a dimming trend of 0.059$\pm 0.005$ mag yr$^{-1}$ in \textit{I}-band, 0.051$\pm 0.005$ mag yr$^{-1}$ in \textit{R}-band and 0.060$\pm 0.005$ mag yr$^{-1}$ in \textit{V}-band (see Figure \ref{img:HH22Lightcurve}).
Hence, the region C, seems to be dimming faster than V1647 Ori. This could be either due to material inflow into cavity between region C and V1647 Ori as the outburst is progressing or due to a change in extinction along the cavity induced by slow dimming of V1647 Ori's brightness.
During the first outburst in 2003, the linear dimming rate during the plateau stage was 0.24 mag yr$^{-1}$ in \textit{R}-band \citep{fedele07}, which was $\sim 6.3$ times faster in magnitude scale than the present dimming rate in second outburst.
Just like in other T-Tauri stars, we also see a lot of short time scale random variations in the source magnitude (peak-to-peak $\Delta V\simeq 0.35$ mag, $\Delta R\simeq 0.30$ mag and $\Delta I\simeq 0.20$ mag), which could be due to density fluctuations in the infalling gas on to the star.

Our lightcurve of V1647 Ori does not show any 56 day periodicity which was reported by \citet{acosta07} during the first 2003 outburst. Based on the correlated reddening of flux during the minima of light curve, they proposed that periodicity was due to occultation of a dense clump in accretion disc at a distance of 0.25 AU from the star. The peak-to-peak amplitude in \textit{I}-band was $\sim$ 0.3 mag in 2003. We have not detected this variability in 2008 outburst which implies the dense clump might have got dissipated between 2003 and 2008 outburst events. Our Lomb-Scargle periodogram analysis of magnitudes did not show any other statistically significant periodicity.

The optical magnitudes during the second outburst are almost similar to that of the first outburst in 2003. In fact the first known outburst of V1647 Ori in 1966 ($\sim$ 38 years prior to 2003), reported by \citet{aspin06}, also had similar magnitude to the present one, however, all these three outbursts have different timescales. Implications of this fact on outburst model will be discussed in Section \ref{modelsNDiscuss}.

\begin{figure}
\begin{center}
\includegraphics[width=1\textwidth]{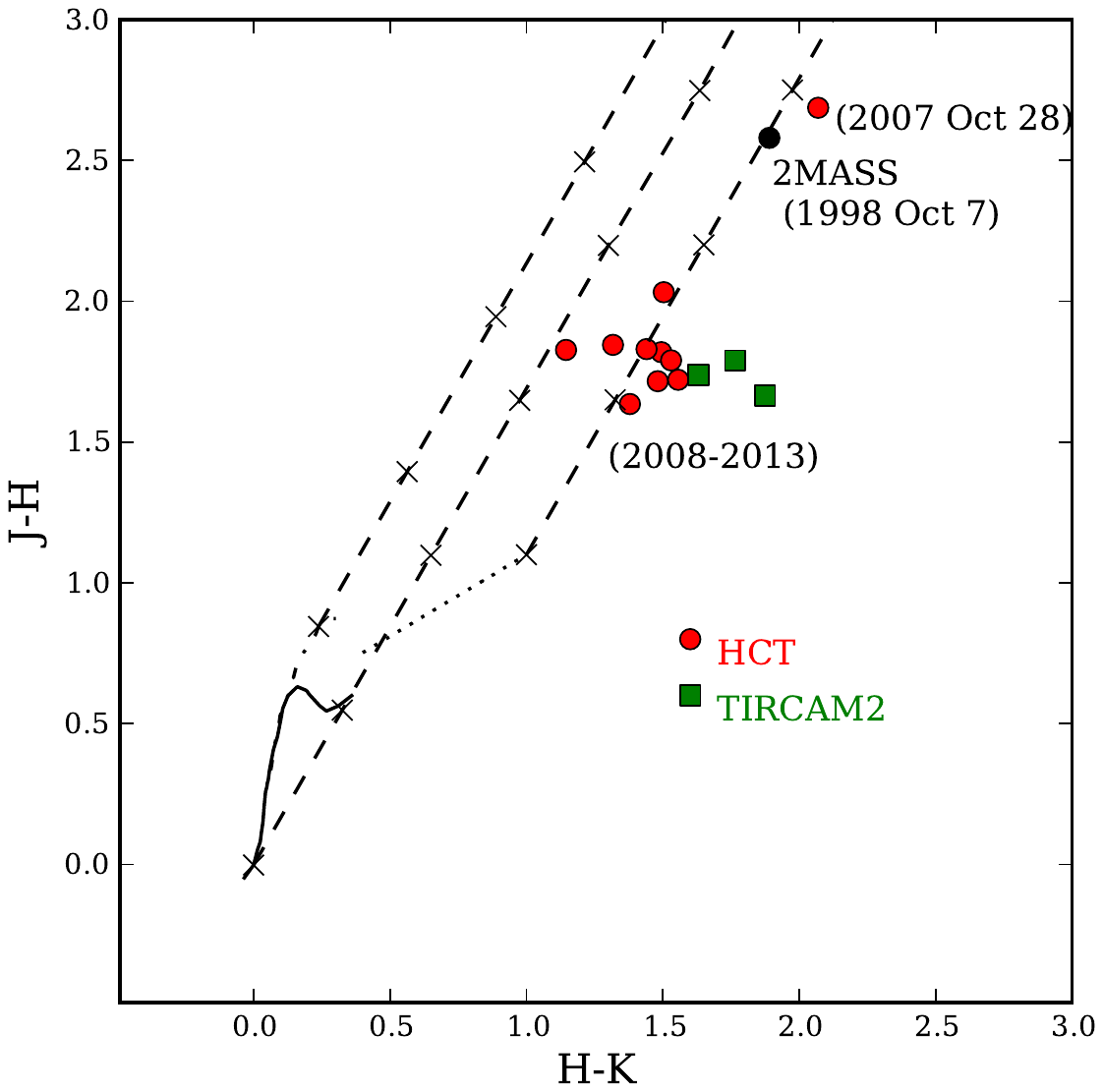}%{JHK_CC.eps}
\caption{Movement of the position of V1647 Ori in $J-H/H-K$ CC diagram from the quiescent phase in 2007 to the second outburst phase. The solid curve shows the locus of field dwarfs and the dashed-dotted curve shows the locus of giants \citep{bessel88}. The dotted line represents the locus of CTT stars \citep{meyer97}. The diagonal straight dashed lines show the reddening vectors \citep{rieke85}, with crosses denoting an $A_{V}$ difference of 5 mag.} 
\label{img:JHK_CCdia}
\end{center}
\end{figure}

Our NIR \textit{J}, \textit{H} and \textit{K} magnitudes are listed in Table \ref{table:JHKmags}. Similar to optical light curve, there is a faint dimming trend in NIR also. \citet{raman13}, with more NIR data points, estimated the fading rate in \textit{J}-band to be 0.08 $\pm 0.02$ mag yr$^{-1}$.
The $J-H/H-K$ color-color (CC) diagram (Figure \ref{img:JHK_CCdia}) shows the movement of V1647 Ori from 2007 data point taken in  quiescent phase to outburst state. It is similar to what was seen in 2003 outburst. 
From the quiescent phase position in CC diagram, V1647 Ori has moved towards the classical T-Tauri (CTT) locus along the redenning vector and presently occupies the same position as in 2003 outburst. The position of V1647 Ori in CC diagram is consistent with similar CC diagram published by \citet{raman13}. This implies that the decrease in line of sight extinction during the outburst is same as that seen during the 2003 outburst. Since our line of sight is through the envelope, it must be likely due to a reversible mechanism like dust sublimation in the inner region of envelope during each outburst \citep{acosta07,mosoni13,aspin09}.  Since the star is deeply embedded, we have reflections and dust emission effects also affecting the position of V1647 Ori in the CC diagram. So the extinction estimated from CC diagram is not very reliable. Otherwise, we can see that the second outburst has cleared out circumstellar matter of $\delta A_{V} \sim 6 \pm 2 $ mag.
This is also consistent with the estimate of extinction change during first outburst by \citet{mosoni13}, $\delta A_{V} \sim 4.5$ mag (see also \citet{aspin08}). %and $\delta A_{I_{c}} \sim 2.5$ mag. 

\subsection{Morphological Results}
\begin{figure}%[h]
\begin{center}
\includegraphics[width=1\textwidth]{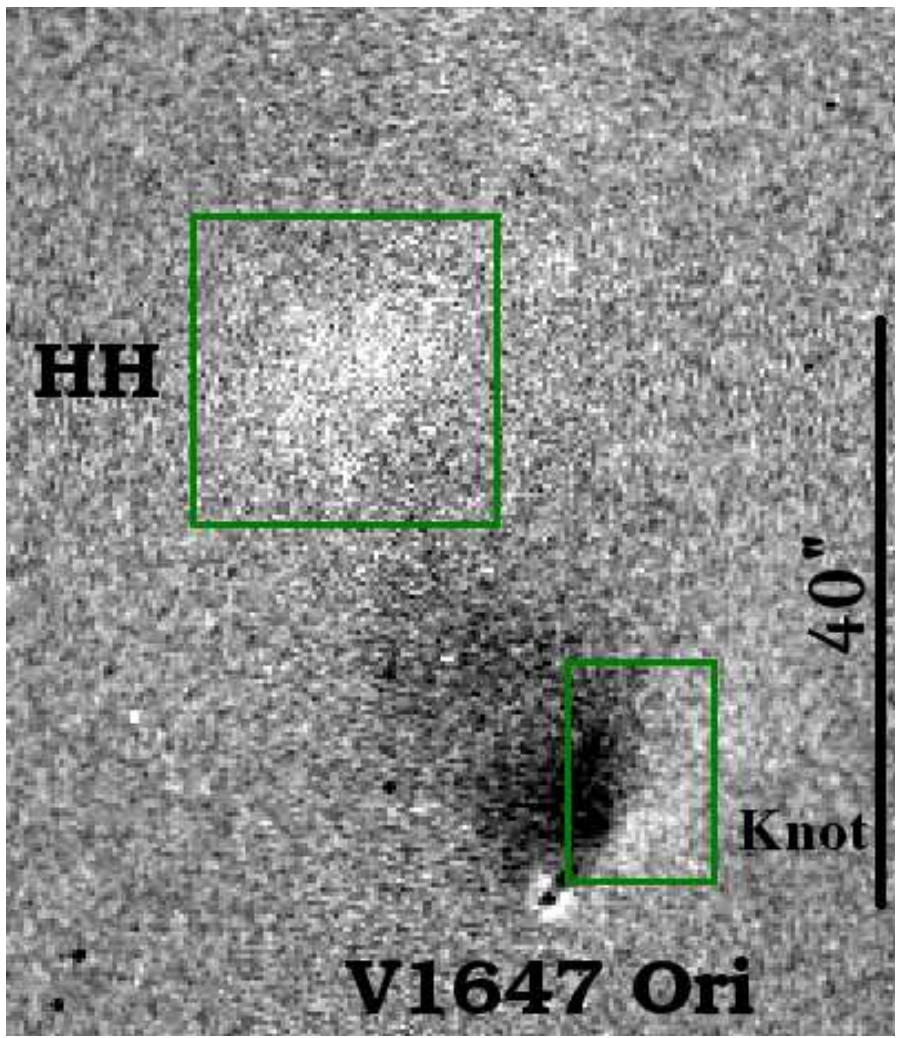}%{R11-R04_mcneil3.eps}
\caption{HCT \textit{R}-band image of 2011 minus 2004, after normalising with respect to the brightness of V1647 Ori. The images were chosen from the nights with same atmospheric seeing and aligned using other field stars in the FoV. The bright portions show the regions which were relatively brighter in 2011 and dark portions show the regions which were relatively brighter in 2004. For example, region C shown in upper box is brighter in 2011. Also note the change in illumination of south-western knot region B marked by lower box in the nebula. North is up and east is to the left-hand side.}
\label{img:R11-R04}
\end{center}
\end{figure}

%-------------------
Between 2003 and 2008 outbursts, the McNeil's nebula does not have any significant morphological changes, however the intensity distribution of the nebula has changed between the outbursts. Figure \ref{img:R11-R04} shows the difference in \textit{R}-band flux along the nebula between 2011 and 2004. Images of similar atmospheric conditions were taken and scaled to match the brightness of V1647 Ori before subtracting 2004 image from that of 2011. Brighter shade implies that region is brighter in 2011 than 2004.  We can see that the region C is brighter in second outburst than it was in 2004. This could be due to dust clearing up between the last two outbursts along the cavity seen in NIR in region C direction \citep{ojha05}. Our photometric results show region C is dimming faster than V1647 Ori and one of the explanations for that could be material inflow into cavity during the outburst. However, region C is relatively brighter in 2008 outburst than in 2004 for the same brightness of V1647 Ori. This implies that the matter inflow to cavity was not occurring during the quiescent phase between 2006 and 2008. 
This is also based on the implicit assumption that the extinction along the line of sight direction to V1647 Ori is same between 2003 and 2008 outbursts. The other significant change is in illumination of the south-western knot (region B) of the nebula; its illumination seems to have shifted slightly towards west. These illumination changes in nebula imply a structural change in the circumstellar matter above the disc and cavity. A similar pattern and conclusion were also reported between 1966, 2003 and 2008 outbursts by \citet{aspin06,aspin09}. Similar analysis of image pairs taken between 2008 and 2012 did not show any significant morphological changes. Our image pairs had a seeing of 1.6\arcsec. Hence, to check  whether any change in illumination of nebula is undergoing during the present outburst, we need images with seeing less than 1.6\arcsec.

\subsection{Spectroscopic Results}
\begin{figure}
  \includegraphics[width=1\textwidth]{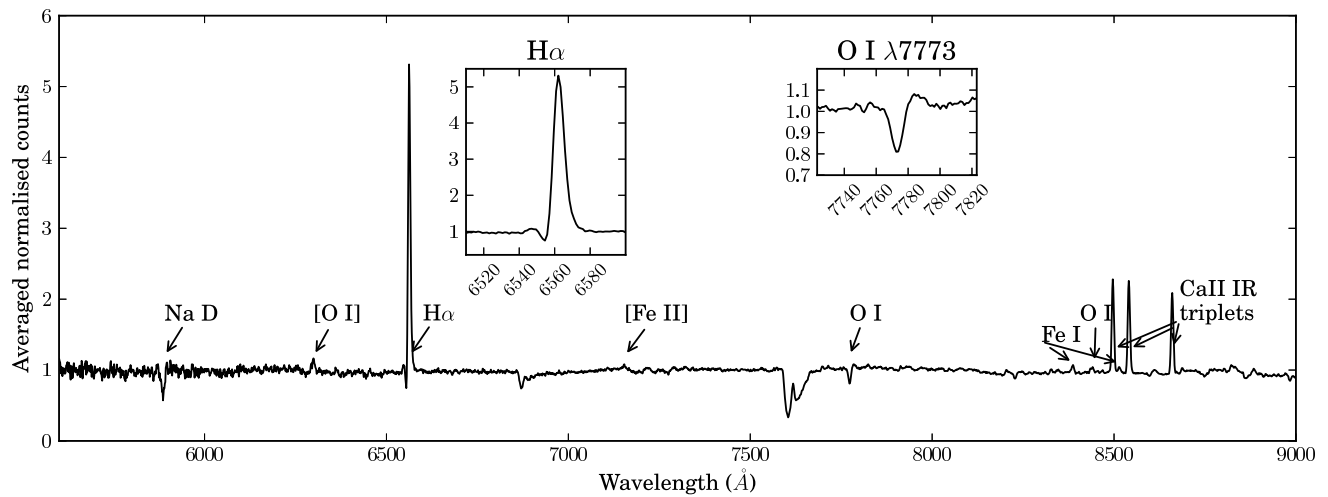}%{Avg_labelled_spectraV1647Ori-withInset.eps}
  \caption{The spectral lines present in the spectrum of V1647 Ori are labelled above. To improve the S/N ratio, the normalised spectrum was obtained by weighted averaging of 33 HCT spectra taken over the outburst period 2008 September to 2013 March,  each with an average exposure time of 40 minutes. The spectra are not corrected for atmospheric absorption lines. The absorption lines which are not labelled are atmospheric lines. H$\alpha$ and OI $\lambda$7773 line profiles are shown more clearly in insets.}
\label{img:V1647Spectra}
\end{figure}

%--------------------
V1647 Ori optical spectra show strong H$\alpha$ (6563 $\mathring{A}$) emission and Ca II IR triplet at 8498, 8542 and 8662 $\mathring{A}$ in emission. Other weak lines seen are Na D (5890+5896 $\mathring{A}$) and OI (7773 $\mathring{A}$) in absorption, and [O I] (6300 $\mathring{A}$), O I (8446 $\mathring{A}$), [Fe II] (7155 $\mathring{A}$) and Fe I (8388, 8514 $\mathring{A}$) in emission %and a few higher level hydrogen Paschen lines in emission 
(see Figure \ref{img:V1647Spectra}). The equivalent widths of H$\alpha$, Ca II IR triplet lines and OI (7773 $\mathring{A}$) are listed in Table \ref{table:eqws}.

\subsubsection{H$\alpha$ line}
\begin{figure}
  \includegraphics[width=1\textwidth]{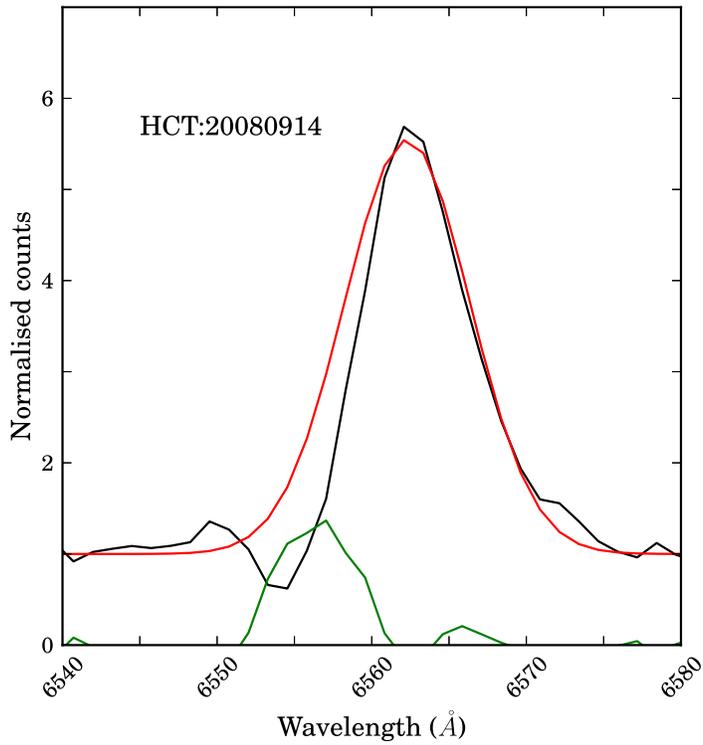}%{HalphaProfile-single.eps}%{Halpha_plots2.eps}
  \caption{The variations of $H\alpha$ profiles during our four and a half year observations. P-Cygni profiles were more prominent in the early part of the outburst in 2008. A Gaussian is fitted to the right wing of the profile in red color and the difference of that Gaussian fit to actual spectra is plotted in green color to see the absorption component clearly. \textit{Figure \ref{img:Halphaplots} with 70 other plots of remaining nights are available in the online version of the Journal.}}
\label{img:Halphaplots}
\end{figure}

\begin{figure}
  \includegraphics[width=1\textwidth]{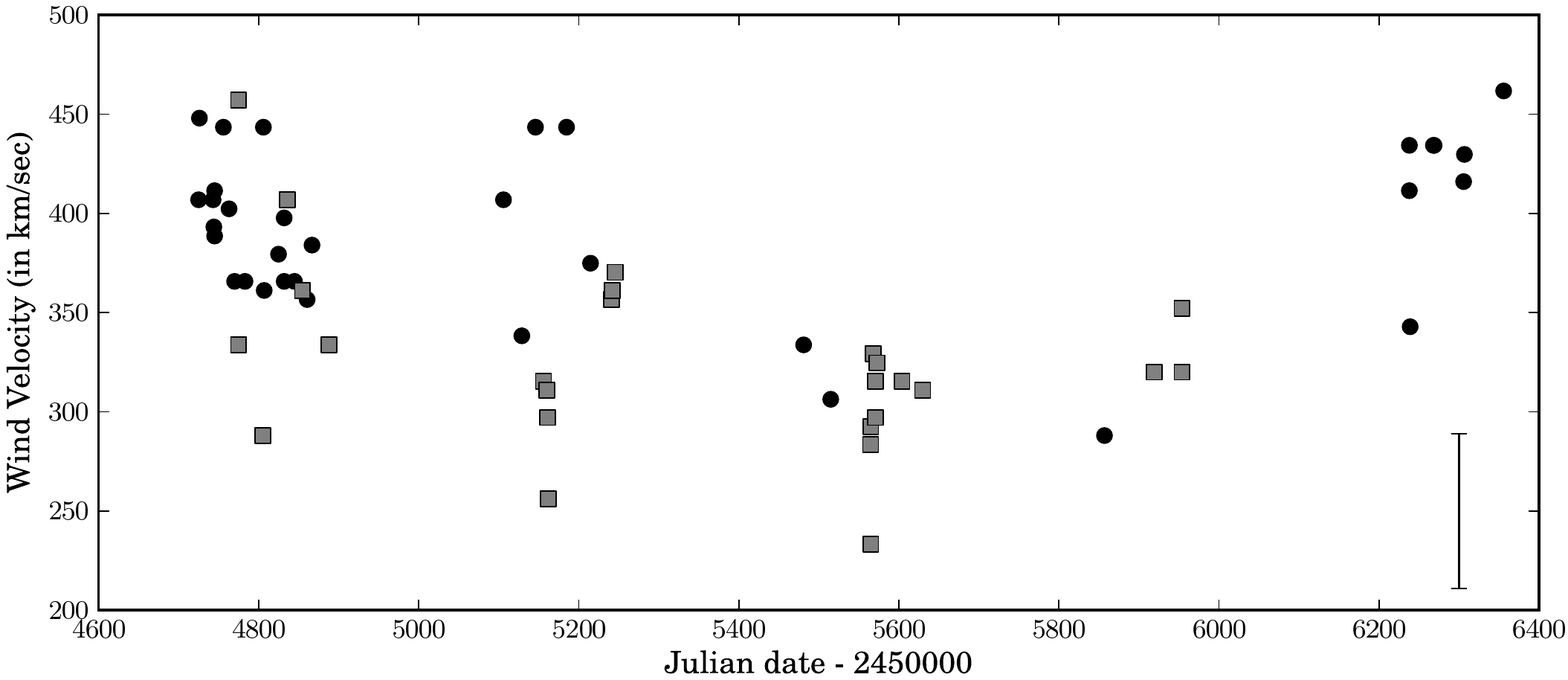}%{JD-WindVelocity.eps}
  \caption{The velocity of expanding wind from the blue-shifted absorption minima in $H\alpha$ P-Cygni profile. The filled circles  are from HCT, and grey filled squares are from IGO measurements. The error bar on the bottom right corner shows the typical $\pm$ 39 km/sec error estimated for data points.}
\label{img:PcygniVelocity}
\end{figure}

\begin{figure}
  \includegraphics[width=1\textwidth]{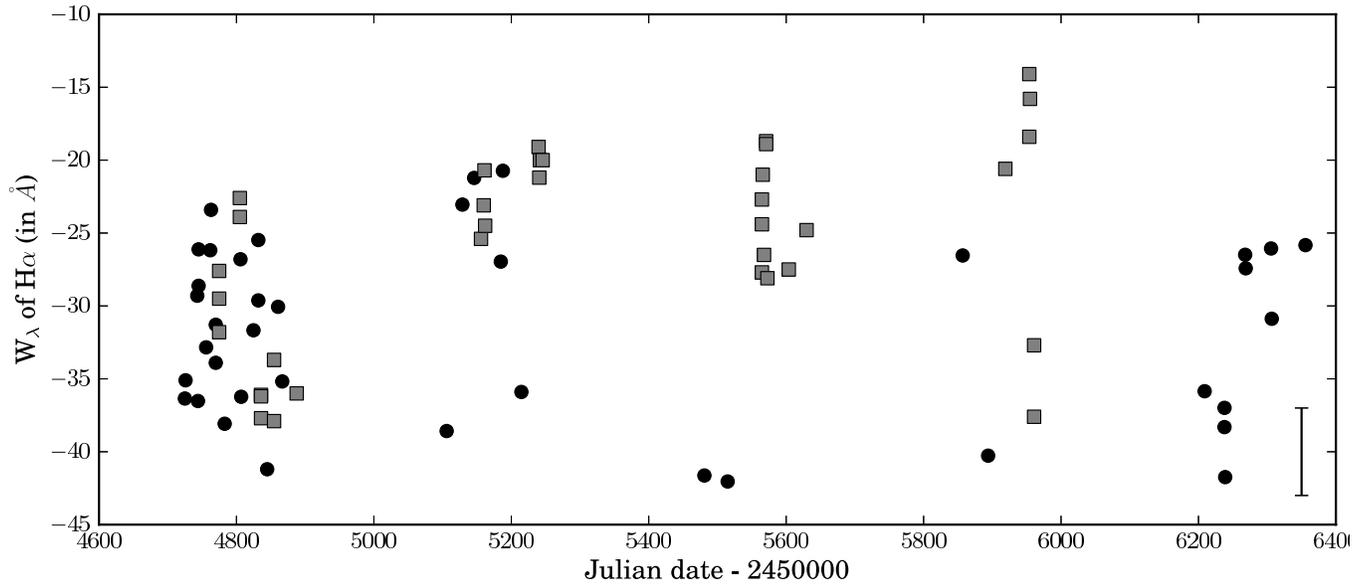}%{JD-Halpha-eqw.eps}
  \caption{ The variation in equivalent width of $H\alpha$ emission. The filled circles are from HCT, and filled grey squares are from IGO measurements. Each data point has an error bar of $\sim \pm 3 \mathring{A}$. This error bar is shown at the right bottom corner.}. 
\label{img:Halpha:EQW}
\end{figure}

Strong $H\alpha$ line in V1647 Ori shows a clear P-Cygni profile as well as substantial variations. Figure \ref{img:Halphaplots} shows the variations of $H\alpha$ profiles during our four and a half years of observations. P-Cygni profiles were more prominent in early part of the outburst in 2008. To see the absorption component clearly, a Gaussian is fitted to the right wing of the profile in red color and the difference of the fit to actual spectra is plotted in green color. Figure \ref{img:PcygniVelocity} shows the outflow velocity and associated error bar of expanding wind from the blue-shifted absorption minima in $H\alpha$ profile. These blue-shifted absorption components were present in 2003 outburst also and disappeared during the fading stage of the outburst \citep{ojha06,fedele07}.  Figure \ref{img:Halpha:EQW} shows the variation in equivalent width ($W_{\lambda}$ in $\mathring{A}$) of $H\alpha$ emission. The calculation of $W_{\lambda}$ was very sensitive to the weak continuum flux around $\lambda6563$ $\mathring{A}$, and the error estimate for each data point is $\sim \pm 3$ $\mathring{A}$. The $W_{\lambda}$ of 2008 outburst is in similar range of that during first outburst in 2003. 
Since $H\alpha$ emission comes from the innermost accretion powered hot zone, we can expect its strength to be proportional to the accretion rate. The optical photometric magnitude in second outburst is similar to that of first outburst, which implies the continuum flux is almost of the same value. Hence from the fact that $W_{\lambda}$ is of similar value as of first outburst, we can deduce that the accretion rate is also of the same order in both the outbursts during its \textquotedblleft high plateau\textquotedblright stage.
 
\subsubsection{Ca II IR triplet lines}
\begin{figure}
  \includegraphics[width=1\textwidth]{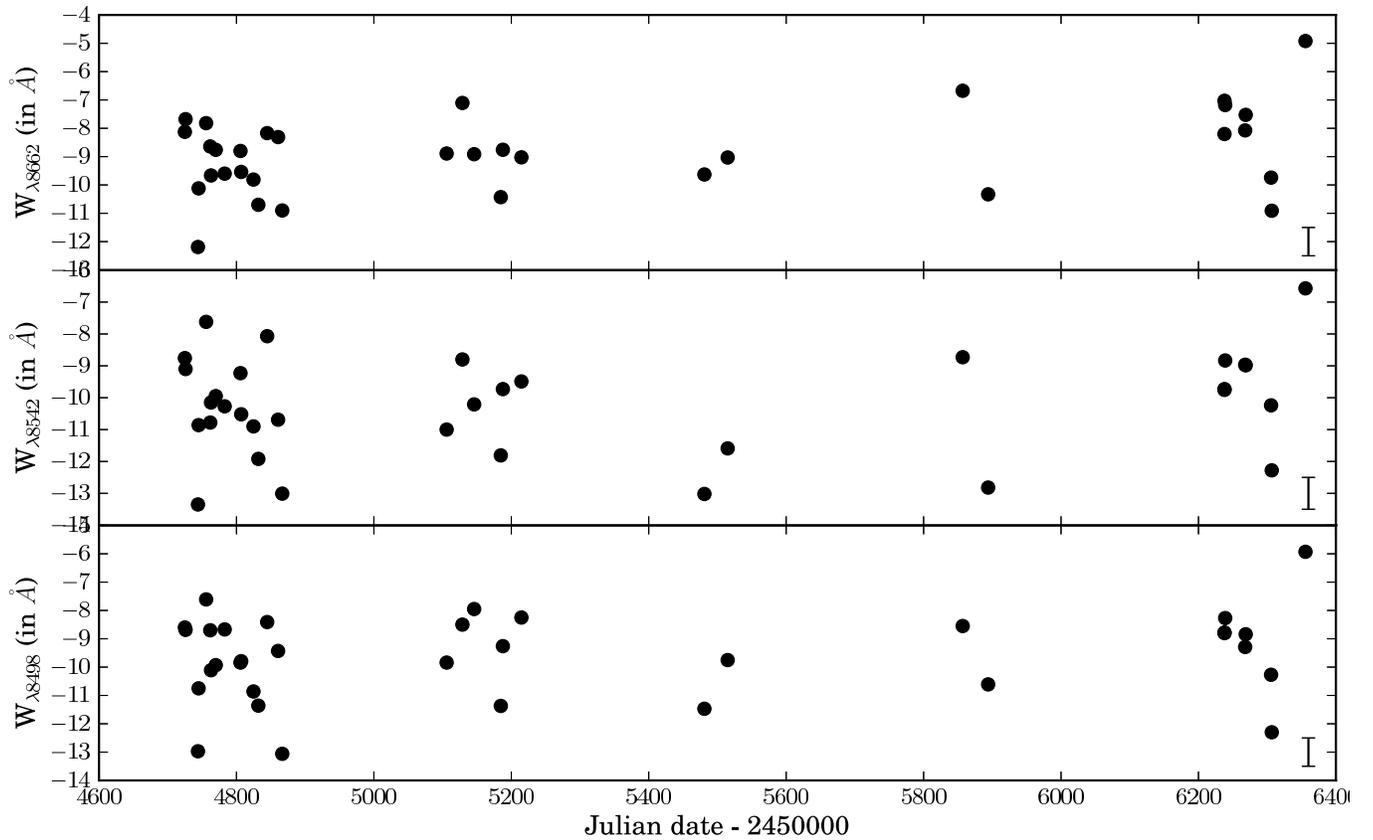}%{JD-CaII_1_2_3_eqw.eps}
  \caption{The variation in equivalent widths of Ca II IR triplet lines (8498, 8542 and 8662 $\mathring{A}$) during the outburst period 2008 September to 2013 March. All data points are from HCT measurement and each point has an error bar of $\sim \pm 0.5 \mathring{A}$. This error bar is shown at the right bottom corner.} Due to limited spectral range in IGO grism, Ca II triplet lines were not observed from IGO.
\label{img:CaIIewq}
\end{figure}

\begin{figure}
  \includegraphics[width=1\textwidth]{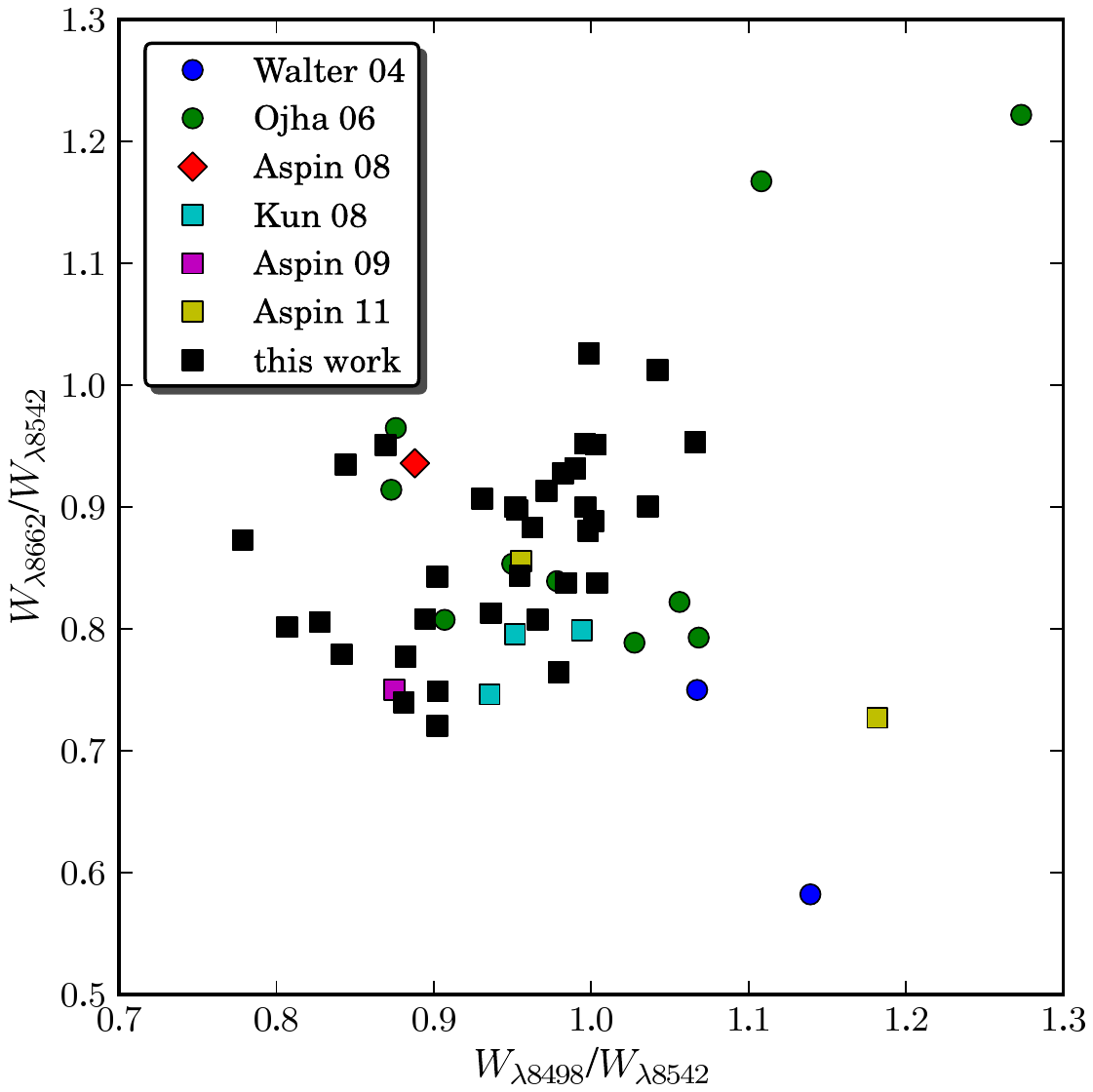}%{CaII-eqwRatio-Scatter.eps}
  \caption{Ratio of Ca II IR triplet lines' equivalent widths, $W_{\lambda8498}$/$W_{\lambda8542}$ versus $W_{\lambda8662}$/$W_{\lambda8542}$, of our data as well as previously published data from literature. 2003 outburst points are shown as circles \citep{walter04,ojha06}, 2008 outburst points are shown as squares \citep[and this work]{kun08,aspin09,aspin11} and 2007 quiescent phase point is shown as diamond \citep{aspin08}.
The typical error bar on our new data (black squares) is $\sim \pm 0.1$.
}
\label{img:CaIIeqwRatios}
\end{figure}

\begin{figure}
  \includegraphics[width=1\textwidth]{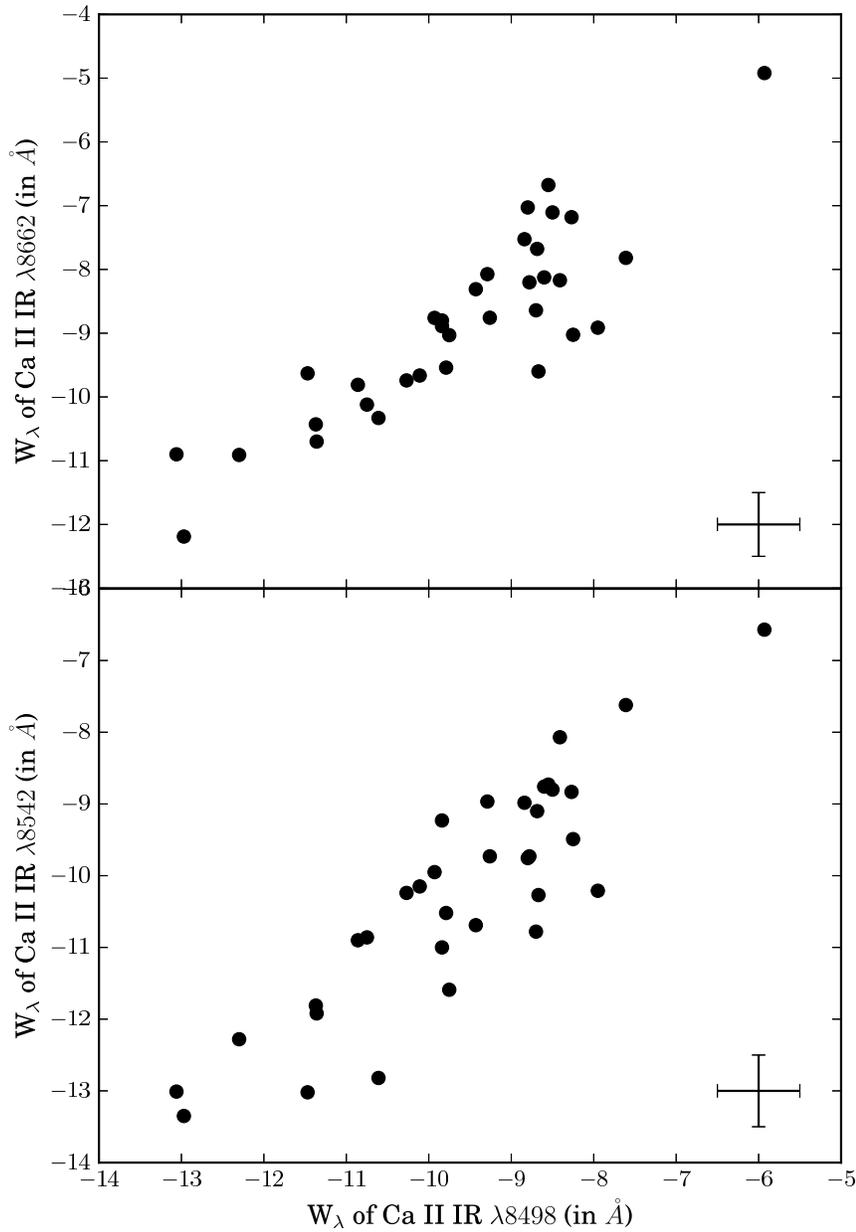}%{CaII_1_2_3_eqwCorr.eps}
  \caption{Strong correlation between the equivalent widths of Ca II IR triplet lines. Typical error bar is given at the right bottom corner. The Pearson correlation coefficient (PCC) between both $W_{\lambda8662}$ and $W_{\lambda8498}$, and $W_{\lambda8542}$ and $W_{\lambda8498}$ is 0.88 with a 2-tailed p value $\ll 0.0001 $.
}
\label{img:correlation:CaII}
\end{figure}

The plots of equivalent widths of Ca II IR triplet lines (8498, 8542 and 8662 $\mathring{A}$) are shown in Figure \ref{img:CaIIewq}. We have much lesser error bars ($\pm 0.5 \mathring{A}$) for the $W_{\lambda}$ due to the higher continuum flux in these wavelengths. The Ca II IR triplet emission lines are seen to be in the ratio $1.07\pm0.09$ : $1.15\pm0.1$ : $1$. This nearly equal ratio is due to optically thick gas with the collision decay rates larger than the effective radiative decay rates of upper states of Ca II lines. Such environment is typically seen in many T-Tauri stars. Optical thickness along with the non-detection of forbidden [Ca II] lines above noise, implies number density of electrons to be $\approx 10^{11}$ cm$^{-3}$ \citep{hamann92}. 
For comparison with Figure 8 of \citet{hamann92}, Figure \ref{img:CaIIeqwRatios} shows the scatter plot between the ratios of equivalent widths, $W_{\lambda8498}$/$W_{\lambda8542}$ and $W_{\lambda8662}$/$W_{\lambda8542}$, of our data as well as previously published data from literature. The error bar in our new data (black squares) is $\pm 0.1$. From the position of 2007 data point (red diamond) taken during quiescent phase \citep{aspin08} in this scatter plot, \citet{aspin08,aspin09} had concluded that the optical density of the region emitting Ca II IR triplet lines changed significantly between outburst and quiescent phases. But since there is only one data point from quiescent phase and it is lying within the scatter of points from ongoing outburst, it might be difficult to conclude that the change in ratios observed was actually due to V1647 Ori moving from quiescent phase to outburst phase. 
We could see strong correlation between the equivalent widths ($W_{\lambda}$) of Ca II IR triplet lines (see Figure \ref{img:correlation:CaII}). 
The Pearson correlation coefficient (PCC) between both $W_{\lambda8662}$ and $W_{\lambda8498}$, and $W_{\lambda8542}$ and $W_{\lambda8498}$ is 0.88 with a 2-tailed p value $\ll 0.0001 $.
This implies that the fluctuations in $W_{\lambda}$ are not random statistical error. It could be due to fluctuations of continuum flux around $\lambda8500 \mathring{A}$. Peak-to-peak fluctuations of $W_{\lambda}$ is $\sim 5 \mathring{A}$, which means if the flux from these lines is assumed to be constant then the continuum flux has fluctuated by a factor of $\sim 1.6$, which in terms of the log scale of magnitude is $\sim 0.5$. This indeed matches with peak-to-peak fluctuation in \textit{I}-band magnitude during the entire period.
We do not see any strong correlation between $W_{\lambda}$ of $H\alpha$ line and Ca II IR triplets. However, it should be noted that the error bars in $W_{\lambda}$ of $H\alpha$ are much higher than that of Ca II IR lines due to low continuum flux around $\lambda6563 \mathring{A}$.

\begin{figure}
  \includegraphics[width=1\textwidth]{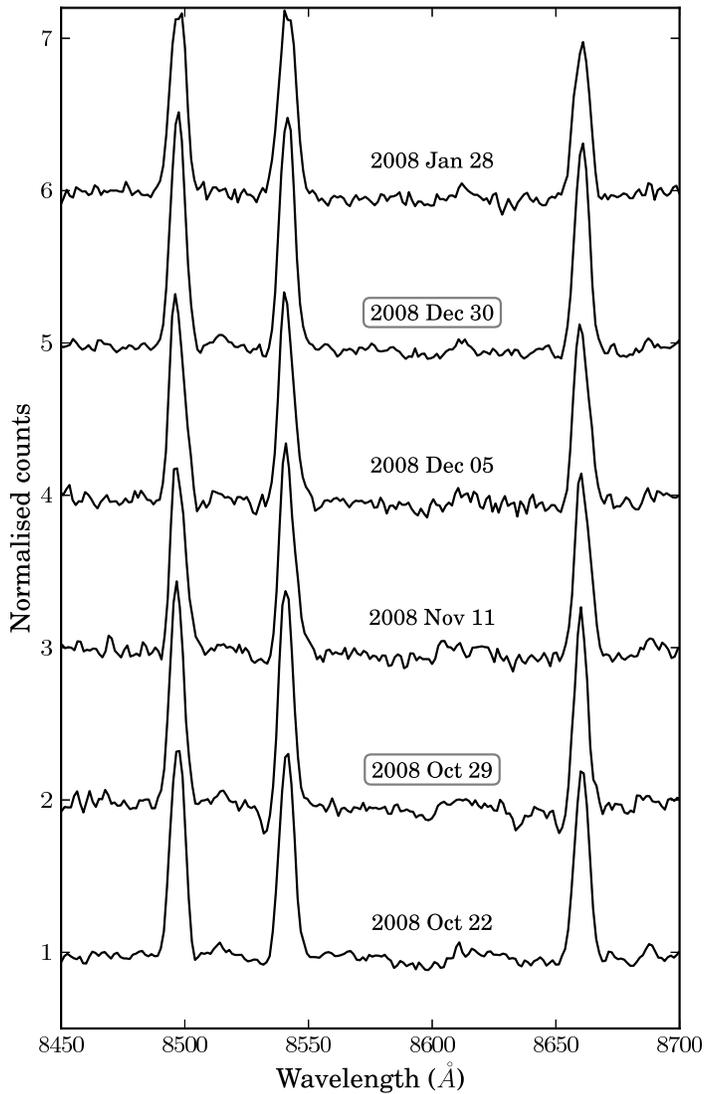}%{CaII_Pcygni_Winter2008.eps}
  \caption{The spectrum of V1647 Ori taken on 2008 October 29 showing clear P-Cygni profile in Ca II IR triplet lines. Similar, but fainter profile was once more detected in 2008 December 30. None of the other nights' spectra showed this profile. For comparison, available nearby nights' spectra are also plotted.}
\label{img:CaII20081029}
\end{figure}

In 2008 October 29 spectrum, we could clearly detect P-Cygni profile in Ca II IR triplet lines (Figure \ref{img:CaII20081029}). The strengths of absorption trough of the three lines were in the same pattern as that of T-Tauri star WL 22 %(see Page: 254 in
 \citep{hamann92}, i.e. the pattern with strongest absorption in $\lambda8542 \mathring{A}$, then $\lambda8662 \mathring{A}$ and very weak in $\lambda8498 \mathring{A}$. The ratio of $W_{\lambda}$ of the blue-shifted absorption between $\lambda8542 \mathring{A}$ and $\lambda8662 \mathring{A}$ is 0.76 : 0.45 =1.69 : 1. This ratio is consistent within error bars to the  1.8 : 1 
%(i.e. 9:5) 
ratio of intensity from atomic transition strength of these lines. Hence, unlike the region producing emission lines, this absorption regions are optically thin. So by using optically thin assumption, we can estimate the column density of Ca II by the formula \citep{spitzerISM78}:\\ 
$ N_{CaII} = 1.1\times10^{20} \times{\frac{\mathring{A}}{\lambda}}^{2} {\frac{1}{f_{lu}}}{\frac{W_{\lambda}}{\mathring{A}}}$ cm$^{-2}$ , \\ 
where the oscillator strength $f_{lu}$ for the lines 8542 and 8662 $\mathring{A}$ are 0.39355 and 0.21478 respectively, taken from \citet{merle11}. Substituting the $W_{\lambda}$, wavelength and oscillator strength for both the lines, we get the column density ($N_{CaII}$) as $2.9\times10^{12}$ cm$^{-2}$ and $3\times10^{12}$ cm$^{-2}$ respectively. This is the column density of Ca II atoms in this small duration of outflow wind. Assuming reasonable estimates of temperature T= 2600K (disc temperature estimated by \citet{rettig05}) and pressure P=1 Pascal (typical pressure in solar winds), we get the fraction of ionisation using Saha's formula to be $\approx 0.004$. Hence, dividing by this fraction we obtain the column density of Ca atoms in gas blob to be  $\approx 7.5 \times10^{14}$ cm$^{-2}$. Assuming solar metalicity, we obtained column density of hydrogen (H) in outflowing gas blob as $\approx 3.4 \times10^{20}$ cm$^{-2}$.
From the doppler shift of the absorption minima, we also obtained the wind velocity to be $313 \pm 10 $ km s$^{-1}$ (in $\lambda8542 \mathring{A}$ line) and $303 \pm 10$ km s$^{-1}$ (in $\lambda8662 \mathring{A}$ line). We also detected a faint P-Cygni profile in $\lambda8542 \mathring{A}$ line on 2008 December 30, with a blue-shifted velocity of $329 \pm 10$ km s$^{-1}$. Apart from these two episodic events, none of our other spectra show any detectable P-Cygni profile. The episodic nature of these two winds implies they are magnetic reconnection driven winds rather than pressure driven steady winds.

\begin{figure}
  \includegraphics[width=1\textwidth]{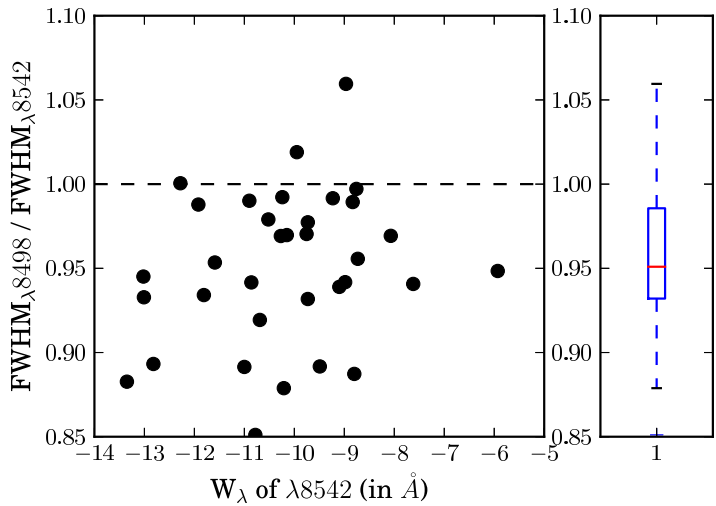}%{RatioCaII_gfwhm1by2-eqw2-wbox.eps}
  \caption{Scatter plot of the ratio of the widths of 8498 and 8542 $\mathring{A}$ versus equivalent width of the line $8542 \mathring{A}$ . Most of the points lie below 1.0 in Y-axis. Box and whisker plot of the distribution is also ploted on the right side. This shows that the $8498 \mathring{A}$ line is slightly narrower than $8542 \mathring{A}$ line.}
\label{img:widthRatio}
\end{figure}

Even though our medium resolution spectra cannot be used to study line widths, since the Ca II IR triplets are near by, we could do relative comparison of the widths of the Ca II IR triplet lines, where width is taken to be the full width half maximum (FWHM) of Gaussian fit of the continuum normalised profile. Since this quantity is the FWHM of Gaussian profile that we get after the convolution of instrument response on the actual line, it is not the FWHM of the line. However, since the lines are very close and the instrumental convolution is common, the wider line will give a wider FWHM after convolution.
Figure \ref{img:widthRatio} shows a scatter plot of the ratio of the widths of $\lambda8498$ $\mathring{A}$ and $\lambda8542$ $\mathring{A}$ versus equivalent width of the line $\lambda8542$ $\mathring{A}$. Most of the points lie below 1.0 in the ratio axis. Using the \textquoteleft test statistic\textquoteright\, for mean with 34 data points, we could reject null hypothesis \textit{$H_{o}$: The mean of the ratio is 1, with 6 sigma confidence.} This shows that the $\lambda8498 \mathring{A}$ line is slightly narrower than $\lambda8542 \mathring{A}$ line. Similar trend is seen in most of the T-Tauri stars \citep{hamann92}. Apart from showing this skewness, since our spectra are of only medium resolution, we cannot quantify the narrowness of the line. This narrowness of optically thinner line $\lambda8498 \mathring{A}$ could be explained by either substantial opacity broadening in $\lambda8542 \mathring{A}$ (since 1:9 is the ratio of atomic line strength) or by lower  dispersion velocity in the inner part of the region of Ca II IR emission \citep{hamann92}.

\begin{figure}
\includegraphics[width=1\textwidth]{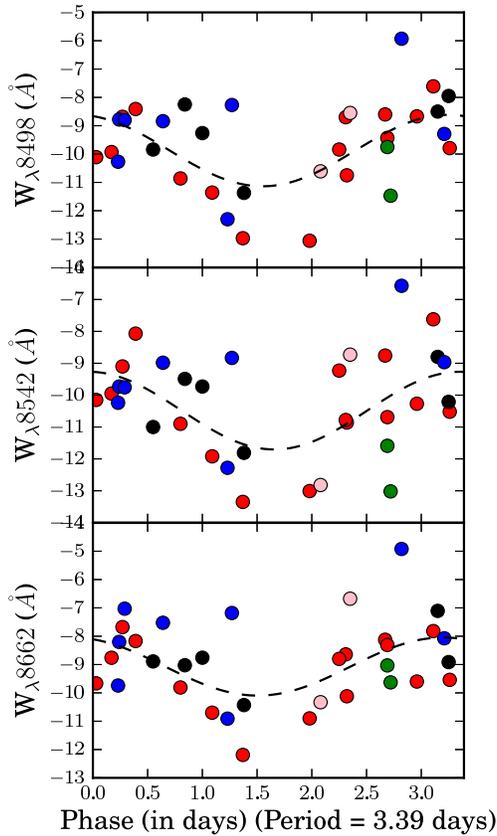}%{Folded_CaII_eqw_3.39.eps}
  \caption{Folded phase plot of the W$_\lambda$ of Ca II IR triplet lines $\lambda$8498, $\lambda$8542 and $\lambda$8662 (in $\mathring{A}$), and \textit{I}-band magnitude. The folding is done over the entire four and a half years of data. The red, black, green, pink and blue circles correspond to data of 2008, 2009, 2010, 2011, and 2012 winter observations respectively. The amplitudes (in $\mathring{A}$) of the three lines in each folded plot are as follows. For a period of 3.39 days:- 1.26 $\pm$ 0.69, 1.22 $\pm$ 0.75, 1.03 $\pm$ 0.67; For a period of 8.09 days:- 1.13 $\pm$ 0.70, 1.33 $\pm$ 0.71, 1.17 $\pm$ 0.65; For a period of 27.94 days:- 1.13 $\pm$ 0.72, 1.11 $\pm$ 0.79, 1.14 $\pm$ 0.67; For a period of 30.8 days:- 1.26 $\pm$ 0.61, 1.25 $\pm$ 0.63, 0.98 $\pm$ 0.58; For a period of 40.77 days:- 1.26 $\pm$ 0.65, 1.42 $\pm$ 0.58, 1.18 $\pm$ 0.59; For a period of 45.81 days:- 0.95 $\pm$ 0.76, 1.29 $\pm$ 0.74, 1.10 $\pm$ 0.68.  \textit{Figures \ref{img:foldedCaIIeqw}.2 - \ref{img:foldedCaIIeqw}.6 are available in the online version of the Journal.}
}
\label{img:foldedCaIIeqw}
\end{figure}

Our period search in the equivalent width ($W_{\lambda}$) of Ca II IR triplet lines found six faint periodicities in all three lines in the range of 1 - 100 days with $\sim$ 2 sigma confidence level in amplitude. The possible periodicities and confidence were estimated by Lomb-Scargle periodogram along with Monte Carlo simulation.
The possible periods are 3.39, 8.09, 27.94, 30.8, 40.77 and 45.81 day. Since the amplitudes are only of 2 sigma confidence level, they can be confirmed only with more observations.
Figure Set \ref{img:foldedCaIIeqw} shows the folded data of the entire four and a half year observations and the statistical significance of the amplitudes. The amplitude of the least square fitted cosine function is $\sim 1$. If this periodicity is due to change in the continuum flux, the corresponding amplitude of magnitude change we expect in logarithmatic \textit{I}-band magnitude is $\sim 0.1$. This is not much above our error in magnitude estimate, so the fact that we do not see similar periodicity in \textit{I}-band magnitude does not rule out the cause of change in $W_{\lambda}$ as change in continuum flux. We also do not see any corresponding  significant periodicity in $W_{\lambda}$  of H$\alpha$.

\citet{aspin08} took the spectrum of V1647 Ori during  the quiescent phase in February 2007. The $W_{\lambda}$ of both Ca II IR  triplet lines and H$\alpha$ were $\sim 3.3$ times the present value. \citet{aspin09b} had used ratio of $W_{\lambda}$ of  H$\alpha$ between 2003 outburst and quiescent phase to estimate change in accretion rate. Similarly, since the continuum flux changed by a factor of $\sim 40 $ between the quiescent and 2008 outburst phase, we can estimate that 
the change in the line flux of both set of lines is by a factor of $\sim 10$. This agrees with the change in accretion rate. Thus the origin of Ca II IR triplet lines are directly connected to the accretion rate just like H$\alpha$. This is in agreement with finding of tight correlation between Ca II IR line flux and accretion rate in T-Tauri stars by \citet{muzerolle98}, which suggests the origin of these lines to be in the magnetospheric infall zone. The similar value of $W_{\lambda}$ of Ca II IR lines with that of 2003 also strengthens the claim that the accretion rate on to the star from inner disc was same during both the outbursts.

\subsubsection{Oxygen lines}
\begin{figure}
  \includegraphics[width=1\textwidth]{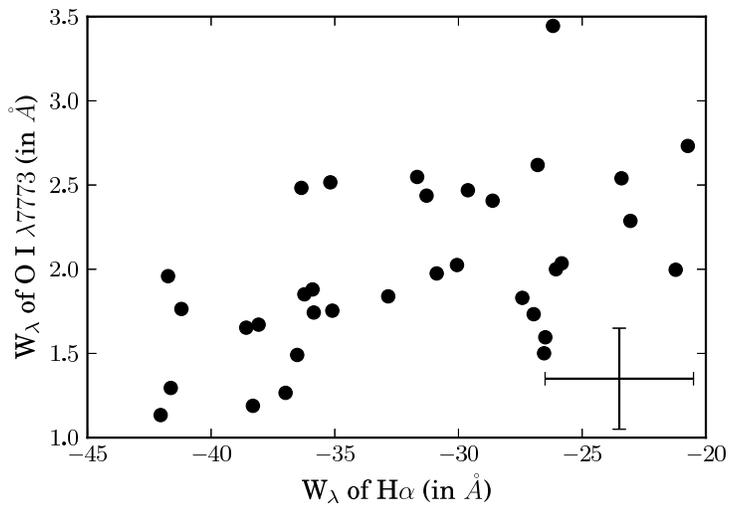}%{Ha_OI_eqw.eps}
  \caption{Correlation between equivalent width of H$\alpha$ and OI $\lambda7773 \mathring{A}$. Typical error bar is given at the right bottom corner. PCC is 0.54 with a 2-tailed p value of 0.001. The weak anti-correlation could be due to a positive correlation between the H$\alpha$ and red-shifted emission component which is filling the absorption component in OI $\lambda7773 \mathring{A}$ profile. Bootstrap analysis gave 95\% confidence range of PCC to be [0.29, 0.72].} 
\label{img:corr_HaVsOI7774}
\end{figure}

The most prominent oxygen line is OI 7773 $\mathring{A}$ in absorption, however, weak OI 8446 $\mathring{A}$ line is also detected in emission. We should be careful in interpreting the $W_{\lambda}$ of OI $\lambda7773$  $\mathring{A}$ absorption line because the profile shape of the line seems to be a combination of red-shifted emission component and more stronger blue-shifted absorption component (see Figure \ref{img:V1647Spectra}). A weak anti-correlation is seen between H$\alpha$ and OI $\lambda7773 \mathring{A}$ (see Figure \ref{img:corr_HaVsOI7774}). Correlation had PCC = 0.54, with a 2 tailed p value of 0.001. This anti-correlation in $W_{\lambda} $ of OI 7773 $\mathring{A}$ and H$\alpha$ could be due to positive correlation between emission component of OI $\lambda7773 \mathring{A}$  filling in the absorption dip and H$\alpha$. Since OI $\lambda7773 \mathring{A}$ cannot be formed in photosphere of cool stars, the absorption component is due to warm gas in the envelope or hot photosphere above disc, while the emission component might be due to the hot gas region from which H$\alpha$ is also being emitted.
\citet{ojha06} reported a decreasing trend in the $W_{\lambda}$ of OI $\lambda7773 \mathring{A}$ from the beginning till end of the 2003 outburst which was interpreted to be possible decrease in turbulence in outer envelope during the outburst period. We do not see such a trend in 2008 outburst, but the values of $W_{\lambda}$ in 2008 outburst remain same as that during the second half of the 2003 outburst.
A slight decrease of $W_{\lambda}$ in the later part of 2003 outburst observed on 2005 September 8, as reported by \citet{ojha06}, in contrast to the increase of $W_{\lambda}$ of other lines, could be due to increase in the $W_{\lambda}$ of OI $\lambda7773 \mathring{A}$ emission component.

\subsubsection{Forbidden lines}
We detected [O I] ($6300 \mathring{A}$) and [Fe II] ($7155 \mathring{A}$) forbidden line emissions in V1647 Ori's spectra. The presence of [O I] $\lambda6300 \mathring{A}$ and [Fe II] $\lambda7155 \mathring{A}$ implies shock regions probably originating from jets. This combined with non-detection of [S II] ($6731 \mathring{A}$) line above our noise level implies the shock region has temperature $T \approx 9000 - 11000 K$  and electron number density $\approx 10^{5} - 10^{6}$ cm$^{-3}$ \citep{hamann94}. We see significant variations in the strengths of the forbidden lines [O I] and [Fe II], however, the lines are too faint in our individual spectra to quantify statistically.
During the fading stage of 2003 outburst in 2006 January, when the bright continuum flux decreased, \citet{fedele07} were also able to detect various strong forbidden line emissions, namely [O I] $\lambda\lambda6300\mathring{A},6363\mathring{A}$, [S II] $\lambda\lambda6717\mathring{A},6731\mathring{A}$ and [Fe II] $\lambda7172\mathring{A}$.

\section{Implication on models of outburst}
\label{modelsNDiscuss}
\begin{figure}
  \includegraphics[width=1\textwidth]{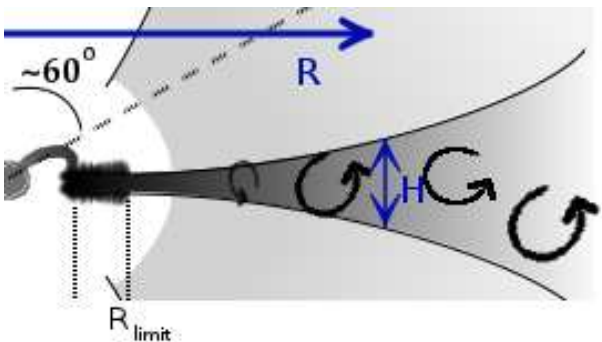}%{Labelled_disc.eps}
  \caption{Cross-section of an $\alpha$ disc is shown above. \textit{R} is the radial distance from the star and \textit{H} is the thickness of the disc, which flares up as the radius \textit{R} increases. $R_{limit}$ is the radius upto which the outburst extends. We are looking into the system at $\sim 60^\circ$ along the dashed line drawn in the figure. In $\alpha$ disc model, the viscosity is due to large turbulent eddies. The speed of eddies is upperbounded by the velocity of sound ($c_{s}$) because any supersonic flow will get dissipated by shock. The size of eddies is also upper bounded by the thickness (\textit{H}) of disc. Thus taking $\alpha$ to be a free parameter $< 1$ we get the viscosity in the disc as $\nu = \alpha c_{s} H$. The orbital velocity is taken to be Keplarian in our problem.}
\label{img:alphadisc}
\end{figure}

The models which are known for FUor / EXor outbursts can be broadly classified into three. 
First kind of model is purely a thermal instability model, initially proposed for dwarf nova systems and was later adapted for FUors kind of outbursts \citep{bell94}. The second kind of model involves a binary companion or planet, which perturbs the disc causing repeated sudden high accretion events \citep{bonnell92,lodato04}. The third kind involves mainly gravitational instability triggering magnetorotational instability (MRI) \citep[and references therein]{zhu09}.
\citet{bell94} model (hereafter BL94) is a pure thermal instability model. In BL94, the thermal instability is triggered when the surface density at a region in disc rises above a critical density. The viscous timescales determine duration of outbursts and quiescent phase. The inner region at a radius $r$ during outburst will deplete below the critical surface density in viscous time scale $\tau_{visc} = r^{2}/{\nu}$. Based on $\alpha$ prescription of viscosity, $\nu = \alpha c_{s} H$, where $c_{s}$ is the isothermal sound speed, $H \approx c_{s}/{\Omega}$ is the disc thickness, $\Omega$ is orbital angular velocity at radius $r$ and $\alpha$ ($< 1$) is a dimensionless parameter (see Figure \ref{img:alphadisc}). Substituting, we get a relation : \\ 
$\tau_{visc} =\frac{r^2 \Omega}{\alpha c_{s}^2} =\frac{1}{\alpha \Omega}{(\frac{r}{H})}^{2}$.

Let $v_{R} \approx {r}/{\tau_{visc}} \approx {\nu}/{r}$ be the effective inward radial velocity component of gas in inner accretion disc. Then the mass infall rate at radius \textit{R} is $ \dot{M}= -2\pi R \Sigma v_{R}$, where $\Sigma$ is the surface density of disc at that radius. Substituting $v_{R}$, $\nu$ and $H$ in above equation of $\dot{M}$, we get $\dot{M} \approx 2\pi \Sigma {\alpha c_{s}^{2}}/{\Omega}$. The time scale of transition between outburst and quiescent phases is much smaller than the viscous timescale. Hence, $\Sigma$ will remain constant, $\Omega$ will also remain constant at a given \textit{R}, which leaves only the parameter $\alpha c_{s}^{2}$ to explain the change by a factor of $\sim$ 10 in inner disc accretion rate to V1647 Ori between the outburst and quiescent phases.
Since the square of isothermal sound speed $c_{s}^{2} = {R_{g} T}/{\mu}$ (where $R_{g}$ is the gas constant, T is temperature and $\mu$ is mean molecular weight), the temperature change by a factor 10 between the phases at the trigger of thermal instability (\citet{zhu09}, Appendix B, their Figure 14) alone will cause a net change in accretion rate by factor of 10. Hence, at least in the inner region of disc, $\alpha$ can change only by a multiplicative factor of order 1 to remain consistent with the observed change in accretion rate.

\begin{figure}
  \includegraphics[width=1\textwidth]{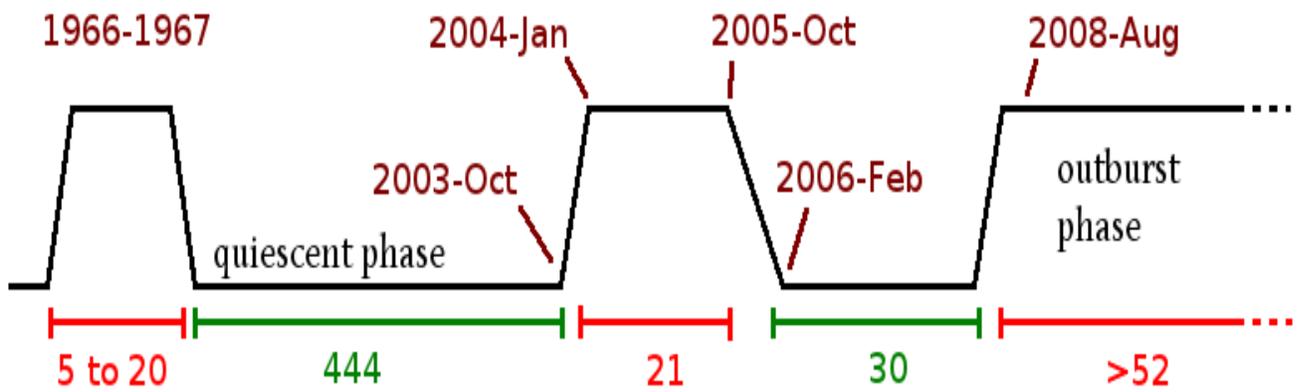}%{lightcurveCartoon3.eps}
  \caption{Optical light curve history of V1647 Ori. The duration of each outburst and quiescent period is marked in units of months. The X-axis of the image is not drawn to scale.}
\label{img:lightcurveCartoon}
\end{figure}

In BL94, assuming the mass infall rate to the inner disc region $ \dot{M}_{in} = constant $, the $\alpha$ determines the timescales and the ratio ${\alpha_{outburst}}/{\alpha_{quiescent}}$ is proportional to the ratio of time in quiescent phase to time in outburst phase (BL94; Table 2). Figure \ref{img:lightcurveCartoon} shows the schematic of the lightcurve of V1647 Ori in optical band we know so far. The photometric magnitudes and overall SED during the quiescent phase in 2007 match with pre 2003 outburst data \citep{aspin08}. Even though accretion rate had fallen by a factor of 10 during quiescent phase, it was still in the order of $10^{-6}$ $M_{\odot} yr^{-1}$. Based on this, \citet{aspin09} had suggested that the 2003 outburst might not have actually terminated in 2006. Since we do not have a good estimate of pre-outburst accretion rate, we shall consider this sudden drop in accretion rate by a factor of 10  as a drop from outburst state to quiescent phase in outburst models. 

Let $\alpha_{o1}$, $\alpha_{o2}$ and $\alpha_{o3}$ represent the three outburst phase $\alpha$ values and $\alpha_{q1}$ and $\alpha_{q2}$ represent the two quiescent phase $\alpha$ values. If we assume $\dot{M}_{in}= constant$ in  BL94, from the ratio of periods, we get the following relations: \\
$\alpha_{o1} = k(22 - 89) \alpha_{q1}$; $\alpha_{o2} =k 21 \alpha_{q1}$; $\alpha_{o3} < k9 \alpha_{q1}$; $\alpha_{o3} <k 0.6 \alpha_{q2}$; where \textit{k} is just the proportionality constant corresponding to the constant $\dot{M}_{in}$.\\
We also get $\alpha_{o2} = (0.23 - 0.95) \alpha_{o1}$; $\alpha_{o3} < 0.4 \alpha_{o2}$; $\alpha_{q2} = 15 \alpha_{q1}$.\\
This is a very huge variation in $\alpha$ parameter. However, our accretion estimates, during the 2003 as well as 2008 outbursts show the accretion rate of matter on to the star from inner disc was quite stable at $\sim$10 times the rate in quiescent phase. Thus the $\alpha$ parameter cannot be fluctuating as much as we estimated; especially $\alpha_{o3} < k 0.6 \alpha_{q2}$ is impossible from the fact that viscosity has to be more in outburst phase than quiescent phase. 
Then the assumption $\dot{M}_{in}= constant$ in BL94 might be wrong. An increase in $\dot{M}_{in}$ can decrease the draining rate of inner disc and can account for a longer duration of outburst period.  It can also explain the slower rate of dimming in the magnitude of the V1647 Ori during its plateau stage in present outburst compared to 2003 outburst.

By letting $\dot{M}_{in}$ to be a variable and substituting $R_{limit}$ to be the radius upto which instability extends, to constrain parameters, we compared the viscous timescale $\tau_{visc} = {R_{limit}^{2}}/{\nu} \approx {R_{limit}^{2} \Omega}/{\alpha c_{s}^{2}}$ between 2003 and 2008 outbursts. The instability triggering temperature at the boundary has to be same for both outbursts, so the sound speed $c_{s}$ will be the same. Parameter $\alpha$ can also be taken to be same during both the outbursts based on our previous conclusion. 
Now the only free variable parameter which determines the timescale is $R_{limit}$. Substituting Keplarian $\Omega \propto R^{-\frac{3}{2}}$, we finally obtain the relation for viscous timescale in terms of $R_{limit}$ to be  
$\tau_{visc} \propto R_{limit}^{\frac{1}{2}}$.

BL94 gives the radius upto which instability extends to be : \\
$ R_{limit} = 20 R_{\odot} {(\frac{\dot{M}_{in}}{3\times10^{-6}M_{\odot}yr^{-1}})}^{\frac{1}{3}} {(\frac{M_{*}}{M_{\odot}})}^{\frac{1}{3}}{(\frac{T_{eff}}{2000})}^{-\frac{4}{3}} \propto {\dot{M}_{in}}^{\frac{1}{3}} $ \\
Substituting this proportionality, we get $\tau_{visc} \propto {\dot{M}_{in}}^{\frac{1}{6}}$. \\
Thus the ratios of mass infall rate from outer to inner disc during each outburst is related to the duration of outburst by the relation  $\frac{\dot{M}_{in_{2008}}}{\dot{M}_{in_{2003}}} > {(\frac{52}{21})}^{6} \approx 230 $. %\\
Thus the infall of gas from outer to inner disc during 2008 outburst is at least 230 times (2 orders) more than that during 2003 outburst. 

Model presented by \citet{zhuI10} includes an MRI instability contribution to viscosity parameter if the temperature of the disc goes above MRI triggering temperature ($T_{M}$) and also includes an effective viscosity contribution from gravitational instability beyond the radius given by Toomre's instability parameter Q. The relation between $R_{limit}$ and $\dot{M}$ in this model is $R_{limit} \propto \dot{M}^{\frac{2}{9}}$ \\
Thus, $\tau_{visc} \propto {\dot{M}_{in}}^{\frac{1}{9}}$,
 and hence $\frac{\dot{M}_{in_{2008}}}{\dot{M}_{in_{2003}}} > {(\frac{52}{21})}^{9}  \approx 3500$.\\
Therefore, the infall of gas from outer to inner disc during 2008 outburst is at least 3500 times (3 orders) more than that during 2003 outburst. 
This estimate is one order more than BL94. It should be kept in mind that the viscosity timescale gives only order of magnitude estimate of the duration of outburst. For a comparison of actual simulated duration of outburst and viscous timescale see Table 1 in \citet{zhuI10}.
If the present outburst continues for a larger period, $\frac{\dot{M}_{in_{2008}}}{\dot{M}_{in_{2003}}}$ ratio will further increase. Since we have an upper limit in $\dot{M}_{in}$ for FUors, the increased ratio could only be explained as a dip in mass inflow in 2003. For example, a gap in disc could have resulted in sudden ending of 2003 outburst. This gap or low density region has to be near the typical $R_{limit}$ predicted by each model, i.e. $\sim 1 AU$. 

To estimate the change in photometric magnitudes between 2003 and 2008 outburst phase due to predicted change in radius $R_{limit}$, we modelled a disc with temperature profile given by outburst accretion rate inside $R_{limit}$ and quiescent accretion rate outside $R_{limit}$. The magnitude variations in optical \textit{I} and NIR \textit{J} bands were found to be less than 1 mag for variation of $R_{limit}$ by a factor of 6. \citet{mosoni13} reported an increase in visibility of resolved interferometric study of V1647 Ori using VLTI/MIDI observations in 8-13 $\mu$m range during the early stage of fading in 2003 outburst (between 2005 March and 2005 September). Apart from the possible scenarios discussed by \citet{mosoni13}, it could also be explained by the relative increase in contribution in total flux from the extended outburst disc when the central star's accretion slowed down. A similar VLTI/MIDI visibility study of the ongoing 2008 outburst will give more input to constrain $R_{limit}$ and outburst models.

Thus, we conclude that a pure thermal instability alone cannot explain the varying timescales of outbursts occurring in V1647 Ori. As proposed for other short rise timescale FUors, V1647 Ori can also be explained only by an outside-in triggering of the instability from outer radius \citep{bell94}. The change in inflow of material from outer to inner disc could be due to many possibilities like MRI, gravitational instability (GI) or planet perturbation.
The smooth surface density assumptions of disc also might not be a good model in light of detection of clump in the disc at 0.27 AU and disappearance of it in second outburst.

Our observations detected a variety of episodic events like sudden short duration winds with hydrogen column density $ \approx 3.4 \times10^{20}$ cm$^{-2}$, fluctuations in H$\alpha$ flux, short timescale variation in continuum flux etc. The short timescale variation in continuum flux could be explained by the convections in the inner disc as suggested by \citet{zhu09} for their model of FU Orionis disc. The variations in $H\alpha$ flux could have origin in some magnetic phenomena in the accretion funnel. The episodic wind events, [Fe II] $\lambda7155$ and [OI] $\lambda6300$ could be originating from jets or disc/stellar winds region.
If we compare between 2008 and 2003 outbursts, the accretion rate on to the star from inner disc, extinction in NIR color-color diagram, outburst magnitude and spectral signatures in optical are the same. The main difference between 2008 compared to 2003 is the larger duration of outburst phase, 6 times slower dimming rate in optical during its \textquotedblleft plateau\textquotedblright\, stage and the change in circumstellar gas distribution revealed by morphological change in nebula's illumination.

\section{Conclusions}
\label{conclusions}
We have carried out four and a half years of continuous monitoring of V1647 Ori in its second outburst phase starting from 2008. Following are our main conclusions:
\begin{enumerate}
 \item V1647 Ori is still in outburst ``plateau`` stage, at similar magnitude to 2003 outburst in optical and NIR bands. It is undergoing a slow dimming at a rate of 0.04 mag yr$^{-1}$, which is 6 times slower than the rate during 2003 outburst. The magnitude shows significant short timescale ($\sim$ 1 day) variations.
\item Morphological studies on illumination of nebula show a consistent change in the circumstellar gas distribution between 2008, 2003 and 1966 outbursts.
\item P-Cygni profiles in H$\alpha$ emission lines show outflowing wind velocities of $\sim$ 350 km s$^{-1}$. Apart from the continuous wind we also detected twice short duration episodic winds driven by magnetic reconnection events, with H column density $\approx 3.4 \times10^{20}$ cm$^{-2}$ from P-Cygni profiles in Ca II IR triplet lines in 2008 October and December. From Ca II IR triplet and H$\alpha$ line strengths, the accretion rate was found to be same as that during the 2003 outburst and is $\sim$ 10 times more than quiescent phase.
\item We could not detect the 56 day periodicity seen in 2003 outburst. 
\item Detection of the forbidden [OI] $\lambda6300$ and [Fe II] $\lambda7155$ lines implies shock regions of $T \approx 9000 - 11000 K$  and $ n_e \approx 10^{5} - 10^{6}$ cm$^{-3}$, probably originating from jets.
\item Timescales of outburst history of V1647 Ori cannot be explained by a simple thermal instability model by \citet{bell94} alone. To explain the large change in accretion rate from outer to inner disc between last two outbursts we require more comprehensive models which includes contribution from MRI, GI and planetary perturbations. From the framework of instability models we conclude that the sudden ending of 2003 outburst could be due to a gap or low density region in inner ($\sim$ 1 AU) disc.
\end{enumerate}
 
\acknowledgments
We thank the anonymous referee for giving us invaluable comments and suggestions that improved the content of the paper.
The authors thank the staff of HCT, operated by Indian Institute of Astrophysics, Bangalore and IGO at Girawali, operated by Inter-University Centre for Astronomy and Astrophysics, Pune for their assistance and support during observations. It is a pleasure to thank J. S. Joshi and all the members of the Infrared Astronomy Group of TIFR for their support during the TIRCAM2 campaign.
All the plots were generated using the 2D graphics environment \textit{Matplotlib} \citep{hunter07}.

\clearpage

\clearpage
\begin{deluxetable}{lcccc}%r}
 
\tabletypesize{\footnotesize}
\tablecolumns{5} 
\tablewidth{0pt}
\tablecaption{Observation log of the photometric and spectroscopic observations \label{table:Obs_Log}}
\tablehead{ \colhead{Date (UT) } & \colhead{JD} & \colhead{FWHM$^\dag$} & \colhead{Filter(s)/grism(s)} & \colhead{ Exposure time (in secs)} }% &\colhead{Notes}}
\startdata
2008 Sep 14 & 2454724 & 1.9\arcsec & $V,R,I,gr8$   & 300, 480, 240, 2400       \\%     & \\
2008 Sep 15 & 2454724 & 1.6\arcsec & $gr8$     & 2400               \\
2008 Sep 16 & 2454726 & 1.5\arcsec & $V,R,I$   & 600, 720, 240       \\%     & \\
2008 Sep 28 & 2454738 & 1.1\arcsec & $V,R,I$ &  240, 240, 180  \\%     & \\
2008 Oct 01 & 2454741 & 1.2\arcsec & $V,R,I$   & 120, 120, 90        \\%     & \\
2008 Oct 02 & 2454742 & 1.2\arcsec & $V,R,I,gr7$ &  240, 240, 180, 2400  \\%     & \\
2008 Oct 03 & 2454743 & 1.5\arcsec & $gr8$     & 2400               \\

\enddata
\tablenotetext{\dag}{Measured average FWHM. This is a measure of the seeing.}
\tablenotetext{\dag\dag}{Observed from IGO, all other nights are from HCT.}
\tablecomments{Table \ref{table:Obs_Log} is published in its entirety in the electronic edition of the Astrophysical Journal. A portion is shown here for guidance regarding its form and content.}
\end{deluxetable}

\begin{deluxetable}{lccccccc}
 
\tabletypesize{\footnotesize}
\tablecolumns{8} 
\tablewidth{0pt}
\tablecaption{Optical $VRI$ photometry of V1647 Ori and region C \label{table:mags}}
\tablehead{ \colhead{Julian Date} & \multicolumn{3}{c}{V1647 Ori} & \colhead{}   &  \multicolumn{3}{c}{region C} \\ 
\cline{2-4} \cline{6-8} \\
\colhead{}  & \colhead{\textit{V}} & \colhead{\textit{R}} & \colhead{\textit{I}} & \colhead{} & \colhead{\textit{V}} & \colhead{\textit{R}} & \colhead{\textit{I}} }
%\hline
\startdata
2454724 & 19.08 & 17.06 & 14.98 & & 15.59 & 14.99 & 14.36 \\
2454726 & 19.11 & 17.11 & 15.04 & & 15.57 & 14.98 & 14.36 \\
2454738 & 19.00 & 17.04 & 14.99 & & 15.59 & 15.06 & 14.36 \\
2454741 & 19.04 & 17.05 & 15.02 & & 15.59 & 14.98 & 14.33 \\
2454742 & 19.09 & 17.09 & 15.04 & & 15.55 & 15.02 & 14.39 \\
2454745 & 19.12 & \nodata & 15.02 & & 15.60 & \nodata & 14.30 \\
2454751 & 19.26 & 17.08 & 15.02 & & 15.50 & 15.02 & 14.20 \\

\enddata
\tablenotetext{\dag\dag}{Observed from IGO, all other nights are from HCT.}
\tablecomments{Estimated error in magnitude of V1647 Ori is $\leq \pm$0.05 for \textit{V} and $\leq \pm$0.02  for \textit{R} and \textit{I}. Estimated error in magnitude of region C is $\leq \pm$0.02 for \textit{V}, \textit{R} and \textit{I}.} 
 
\tablecomments{Table \ref{table:mags} is published in its entirety in the electronic edition of the Astrophysical Journal. A portion is shown here for guidance regarding its form and content.}

\end{deluxetable}

\begin{deluxetable}{lcccr}
 
\tabletypesize{\footnotesize}
\tablecolumns{5} 
\tablewidth{0pt}
\tablecaption{  NIR $J H K$ photometry of V1647 Ori \label{table:JHKmags}}
\tablehead{ \colhead{Date (UT)} & \colhead{\textit{J}} & \colhead{\textit{H}} & \colhead{\textit{K}} & \colhead{Instrument}}
%\hline
\startdata
28-Oct-07 & 14.12 & 11.43 & 9.36 & NIRCAM \\
19-Oct-08 & 10.64 & 9.01 & 7.63 & NIRCAM \\
12-Jan-09 & 10.59 & 8.76 & 7.62 & NIRCAM \\
12-Feb-09 & 11.04 & 9.01 & 7.51 & NIRCAM \\
18-Feb-09 & 10.87 & 9.16 & 7.68 & NIRCAM \\
23-Oct-09 & 10.71 & 8.89 & 7.40 & NIRCAM \\
24-Oct-09 & 10.71 & 8.88 & 7.44 & NIRCAM \\
20-Feb-10 & 10.75 & 8.96 & 7.43 & NIRCAM \\
23-Mar-10 & 10.73 & 9.14 & \nodata & NIRCAM \\
18-Nov-11 & 10.75 & 8.90 & 7.58 & NIRCAM \\
03-Dec-11 & 10.94 & 9.28 & 7.41 & TIRCAM2 \\
04-Dec-11 & 10.97 & 9.19 & 7.42 & TIRCAM2 \\
06-Dec-11 & 10.84 & 9.10 & 7.47 & TIRCAM2 \\
07-Nov-12 & 10.73 & 9.01 & 7.46 & NIRCAM \\

\enddata
\tablecomments{Estimated error in magnitude is $\leq \pm$0.1 (\textit{K}) and $\leq \pm$0.05 (\textit{H} and \textit{J}).}

\end{deluxetable}

\begin{deluxetable}{lccccc}
 
\tabletypesize{\footnotesize}
\tablecolumns{5} 
\tablewidth{0pt}
\tablecaption{Equivalent widths (in $\AA$) of optical lines in V1647 Ori \label{table:eqws}}
\tablehead{ \colhead{JD} & \colhead{H$\alpha$} & \colhead{Ca II $\lambda$8498} & \colhead{Ca II $\lambda$8542}  & \colhead{Ca II $\lambda$8662} & \colhead{OI $\lambda$7773} }
\startdata
2454724  & -36.35  &   -8.601   & -8.758    &  -8.125   &  2.483    \\
2454725  & -35.1   &   -8.686   & -9.1      &  -7.678   &  1.754    \\
2454743  & -36.52  &   -12.97   & -13.35    &  -12.19   &  1.491    \\
2454744  & -28.62  &   -10.75   & -10.86    &  -10.12   &  2.407    \\
2454755  & -32.84  &   -7.61    & -7.621    &  -7.819   &  1.839    \\
%............many more............
\enddata
\tablenotetext{\dag\dag}{Observed from IGO, all other nights are from HCT.}
\tablecomments{Estimated error in equivalent widths of H$\alpha$ lines is $\sim \pm$3 $\AA$, error for Ca II IR triplet lines is $\sim \pm$0.5 $\AA$ and the error for O I $\lambda$7773 lines is $\sim \pm$0.3 $\AA$.} %\\ 
\tablecomments{Table \ref{table:eqws} is published in its entirety in the electronic edition of the Astrophysical Journal. A portion is shown here for guidance regarding its form and content.}

\end{deluxetable}


\begin{thebibliography}{}
\bibitem[\'{A}brah\'{a}m et al., 2006]{abraham06} \'{A}brah\'{a}m, P., Mosoni, L., Henning, T., et al. 2006, A\&A, 449, 13
\bibitem[Acosta-Pulido et al., 2007]{acosta07} Acosta-Pulido, J. A., Kun, M., \'{A}brah\'{a}m, P., et al. 2007, AJ, 133, 2020
\bibitem[Andrews et al., 2004]{andrews04} Andrews, S. M., Rothberg, B. \& Simon, T. 2004, ApJL, 610, L45
\bibitem[Aspin et al., 2006]{aspin06} Aspin, C., Barbieri, C., Boschi, F., et al. 2006, AJ, 132, 1298
\bibitem[Aspin et al., 2008]{aspin08} Aspin, C., Beck, T. L. \& Reipurth, B. 2008, AJ, 135, 423-440
\bibitem[Aspin \& Reipurth, 2009]{aspin09b} Aspin, C. \& Reipurth, B. 2009, AJ, 138, 1137
\bibitem[Aspin et al., 2009]{aspin09} Aspin, C., Reipurth, B., Beck, T. L., et al. 2009, ApJL, 692, L67
\bibitem[Aspin, 2011]{aspin11} Aspin, C. 2011, AJ, 142, 135
\bibitem[Bell \& Lin, 1994]{bell94} Bell, K.R. \& Lin, D.N.C. 1994, ApJ, 427, 987
\bibitem[Bessell \& Brett, 1988]{bessel88} Bessell, M. S. \& Brett, J. M. 1988, PASP, 100, 1134
\bibitem[Bonnell \& Bastien, 1992]{bonnell92} Bonnell, I. \& Bastien, P. 1992, ApJ, 401, 654
\bibitem[Brice\~{n}o et al., 2004]{briceno04} Brice\~{n}o, C., Vivas, A. K., Hern\'{a}ndez, J., et al. 2004, ApJ, 606, L123
\bibitem[Evans et al., 2009]{evans09} Evans II, N. J., Dunham, M. M., J{\o}rgensen, J. K., et al. 2009, ApJS, 181, 321
\bibitem[Fedele et al., 2007]{fedele07} Fedele, D., Van den Ancker, M. E., Petr-Gotzens, M. G. \& Rafanelli, P. 2007, A\&A, 472, 207
\bibitem[Hamaguchi et al., 2012]{hamaguchi12} Hamaguchi, K., Grosso, N., Kastner, J. H., et al. 2012, ApJ, 754, 32
\bibitem[Hamann \& Persson, 1992]{hamann92} Hamann, F. \& Persson, S. E. 1992, ApJS, 82, 247
\bibitem[Hamann, 1994]{hamann94}Hamann, F. 1994, ApJS, 93, 485
\bibitem[Hartmann, 1998]{hartmann98} Hartmann, L. 1998, Accretion Processes in Star Formation, Cambridge Univ. Press, Cambridge
\bibitem[Hartmann \& Kenyon, 1996]{hartmann96} Hartmann, L. \& Kenyon, S. J. 1996, ARA\&A, 34, 207
\bibitem[Herbig, 1977]{herbig77} Herbig, G. H. 1977, ApJ, 217, 693
\bibitem[Hunt et al, 1998]{hunt98} Hunt, L. K., Mannucci, F., Testi, L., et al. 1998, AJ, 115, 2594
\bibitem[Hunter, 2007]{hunter07} Hunter, J. D. 2007, Computing in Science \& Engineering, 9, 90
\bibitem[Ioannidis \& Froebrich, 2012]{ioannidis12} Ioannidis, G. \& Froebrich, D. 2012, MNRAS, 425, 1380
\bibitem[Kastner et al., 2004]{kastner04} Kastner, J. H., Richmond, M., Grosso, N., et al. 2004, Nature, 430, 429
\bibitem[Kenyon et al., 1990]{kenyon90} Kenyon, S. J., Hartmann, L. W., Strom, K. M. \& Strom, S. E. 1990, AJ, 99, 869
\bibitem[K\'{o}sp\'{a}l et al., 2005]{kospal05} K\'{o}sp\'{a}l, A., \'{A}brah\'{a}m, P., Acosta-Pulido, J., et al. 2005, Inf. Bull. Var. Stars, 5661,1
\bibitem[Kun, 2008]{kun08} Kun, M. 2008, Information Bulletin on Variable Stars, 5850, 1
\bibitem[Landolt, 1992]{landolt92} Landolt, A.U. 1992, AJ, 104, 340
\bibitem[Lodato \& Clarke, 2004 ]{lodato04} Lodato, G. \& Clarke, C.J. 2004, MNRAS, 353, 841
\bibitem[McNeil, 2004]{mcneil04} McNeil, J.W. 2004, IAU Circ. 8284
\bibitem[Merle et al., 2011]{merle11} Merle, T., Th\'{e}venin, F., Pichon, B. \& Bigot, L. 2011, MNRAS, 418, 863
\bibitem[Meyer et al., 1997]{meyer97}Meyer, M. R., Calvet, N., \& Hillenbrand, L. A. 1997, AJ, 114, 288
\bibitem[Mosoni et al., 2013]{mosoni13} Mosoni, L., Sipos, N., \'{A}brah\'{a}m, P., et al. 2013, A\&A, 552, A62
\bibitem[Muzerolle et al., 1998]{muzerolle98} Muzerolle, J., Hartmann, L. \& Calvet, N. 1998, AJ, 116, 455
\bibitem[Naik et al., 2012]{naik12} Naik, M. B., Ojha, D. K., Ghosh, S. K., et al. 2012, Bull. Astr. Soc. India, 40, 531
\bibitem[Ninan et al., 2012]{ninan12}Ninan, J. P., Ojha, D. K., Mallick, K. K., et al. 2012, CBET, 3164, 1
%\bibitem[Ninan et al., 2011]{ninan11} Ninan J.P., et al, 2011 ASInC, 4, 1
\bibitem[Ojha et al., 2006]{ojha06} Ojha, D. K., Ghosh, S. K., Tej, A., et al. 2006, MNRAS, 368, 825
\bibitem[Ojha et al., 2005]{ojha05}  Ojha, D. K., Kusakabe, N., Tamura, M., et al. 2005, PASJ, 57, 203
\bibitem[Pringle, 1981]{pringle81} Pringle, J. E. 1981, ARA\&A, 19, 137
\bibitem[Reipurth \& Aspin, 2004]{reipurth04} Reipurth, B. \& Aspin, C. 2004, ApJL, 606, L119
\bibitem[Rettig et al., 2005]{rettig05} Rettig, T. W., Brittain, S. D., Gibb, E. L., Simon, T. \& Kulesa, C. 2005, ApJ, 626, 245
\bibitem[Rieke \& Lebofsky, 1985]{rieke85} Rieke, G. H. \& Lebofsky, M. J. 1985, ApJ, 288, 618
\bibitem[Scholz et al., 2013]{scholz13} Scholz, A., Froebrich, D. \& Wood, K. 2013, MNRAS, 430, 2910
\bibitem[Spitzer, 1978]{spitzerISM78} Spitzer Jr, L. 1978, Physical Processes in the Interstellar Medium (New York:  Wiley)
\bibitem[Teets et al., 2011]{teets11} Teets, W. K., Weintraub, D. A., Grosso, N., et al. 2011, ApJ, 741, 83
\bibitem[Vacca et al., 2004]{vacca04} Vacca, W. D., Cushing, M. C. \& Simon, T. 2004, ApJL, 609, L29
\bibitem[Venkata Raman et al., 2013]{raman13} Venkata Raman, V., Anandarao, B. G., Janardhan, P. \& Pandey, R. 2013, RAA, in press (arXiv:1301.2110)
\bibitem[Vig et al., 2006]{vig05} Vig, S., Ghosh, S. K., Kulkarni, V. K. \& Ojha, D. K. 2006, A\&A, 446, 1021
\bibitem[Walter et al., 2004]{walter04} Walter, F. M., Stringfellow, G. S., Sherry, W. H. \& Field-Pollatou, A. 2004, AJ, 128, 1872
\bibitem[Zhu et al., 2009]{zhu09} Zhu, Z., Hartmann, L., Gammie, C. \& McKinney, J. C. 2009, ApJ, 701, 620-634
\bibitem[Zhu et al., 2010]{zhuI10} Zhu, Z., Hartmann, L., Gammie, C. F., et al. 2010, ApJ, 713, 1134-1142

\end{thebibliography}
\end{document}